\documentclass[aps,showpacs,showkeys, preprint,superscriptaddress]{revtex4}

\usepackage{amsfonts}
\usepackage{amsmath}
\usepackage{amssymb}
\usepackage{latexsym}
\usepackage{times}
\usepackage{epsfig}
\usepackage{multirow}
\usepackage{hyperref}
\usepackage{url}
\usepackage{graphicx}
\input{Qcircuit}

\newcommand\beq{\begin{equation}}
\newcommand\eeq{\end{equation}}
\newcommand\bea{\begin{eqnarray}}
\newcommand\eea{\end{eqnarray}}

\newtheorem{lemma}{Lemma}

 \def\squarebox#1{\hbox
to #1{\hfill\vbox to #1{\vfill}}}
\def\qed{\hspace*{\fill}\vbox{\hrule\hbox{\vrule\squarebox{.667em}\vrule}\hrule}
} 
\newenvironment{proof}{\begin{trivlist}\item[]{\bf Proof:}}{\qed
\end{trivlist}}

\begin{document}

\title{Local Fault-Tolerant Quantum Computation}

\author{Krysta M. Svore}
\email{kmsvore@cs.columbia.edu}
\affiliation{Columbia University, 1214 Amsterdam Ave. MC:0401, New
York, NY 10025}

\author{Barbara M. Terhal}
\email{terhal@watson.ibm.com}
\affiliation{IBM Watson Research Center, P.O. Box 218, Yorktown
Heights, NY 10598, USA}

\author{David P. DiVincenzo}
\email{divince@watson.ibm.com}
\affiliation{IBM Watson Research Center, P.O. Box 218, Yorktown
Heights, NY 10598, USA}

\keywords{quantum computation, fault-tolerance, locality}
\date{}

\pacs{03.67.Lx, 03.67.Pp}

\begin{abstract}
We analyze and study the effects of locality on the
fault-tolerance threshold for quantum computation. We analytically
estimate how the threshold will depend on a scale parameter $r$
which characterizes the scale-up in the size of the circuit due to
encoding. We carry out a detailed semi-numerical threshold
analysis for concatenated coding using the 7-qubit CSS code in the
local and the `nonlocal' setting. First, we find that the
threshold in the local model for the $[[7,1,3]]$ code has a $1/r$
dependence, which is in correspondence with our analytical
estimate. Second, the threshold, beyond the $1/r$ dependence, does
not depend too strongly on the noise levels for transporting
qubits. Beyond these results, we find that it is important to look
at more than one level of concatenation in order to estimate the
threshold and that it may be beneficial in certain places, like in
the transportation of qubits, to do error correction only
infrequently.
\end{abstract}

\maketitle

\tableofcontents



\section{Introduction}

The issue of fault-tolerance is central to the future of quantum
computation. Most studies of fault-tolerance until now
\cite{AB:faulttol,KLZ:faulttol,preskill:faulttol,TB:ft} have
focused on deriving fault-tolerance in a setting where gates
between any two qubits can be executed instantaneously, i.e.\
without taking into account the potential necessity to move qubits
close together in space prior to gate execution. We call this
setting
 the nonlocal model. Current estimates of the fault-tolerance threshold in the probabilistic independent nonlocal error model
can be found in the extensive studies performed by Steane
\cite{steane:overhead}, estimating the threshold failure
probability as $O(10^{-3})$. The recent results by Knill
\cite{knill:ftps} and Reichardt \cite{reichardt:ft} even give
estimates that can be an order of magnitude better, i.e.\
$O(10^{-2})$.

It has been argued, see
\cite{AB:faulttol,steane:overhead,steane:arch} and the analysis in
\cite{gottesman:localft}, that the local model, where qubit
transportation is required, would still allow for a
fault-tolerance threshold, albeit somewhat lower than in the
nonlocal model. However, there has not been any assessment of how
exactly locality influences the threshold, i.e.\ what is the
dependence on the code, the spatial size of the error correction
procedure, the failure rates on the qubit wires, etc. Such an
assessment is timely, because the post-selected schemes by Knill
\cite{knill:ftps} in which large entangled states are prepared in
a trial-and-error fashion (and to a smaller certain extent also
the ancilla preparation procedure proposed by Reichardt
\cite{reichardt:ft}) may fare worse compared to the more
`conventional' methods of computation and error correction when
locality is taken into account. This is because the method of
post-selection is based on attempting to create many states in
parallel, of which a few may pass the test and are integrated in
the computation. If the success probability is low, then at no
additional cost in the nonlocal model, one can increase the number
of parallel tries of creating these states. In the local model,
however, it must be taken into account that an increase in the
number of parallel tries increases the amount of qubit movement,
and thus the potential for errors.

In the first part of this paper, we make a purely analytical
estimate of the threshold when locality is taken into account and
show its dependence on a scale factor $r$, which is a measure of
the spatial scale-up that is due to coding.  This estimate can be
applied to all known error models for which a fault-tolerance
threshold result currently exists.

Since this estimate may be very rough, we set out in the second
part of this paper to analyze and compare, using the
`conventional' method of error correction as described by Steane
in \cite{steane:overhead}, the fault-tolerant behavior for the
concatenated 7-qubit CSS $[[7,1,3]]$ code for the local and
nonlocal model.


In our analysis, we focus on concatenated coding and the threshold
result.  This is not to say that the strategy of using a large
code once so that logical failure rates are small enough for the
type of computation that we envision (see \cite{steane:nature})
may not be of equal or greater practical interest. In such a
scenario, one `merely' has to optimize the error correction
procedures and encoded gate operations for locality.

Here are some of our semi-analytical findings for the 7-qubit code. In these studies
we have used the nonlocal error correction routine and have looked at the effects
of the noise level during transportation of qubits and the scale-up of the computation
due to coding.
\begin{itemize}
\item In the entirely nonlocal setting, we find that one really
needs to look at higher levels of concatenation to estimate a
correct threshold. For the model where all gates have the same
failure probability $\gamma_{else}$ and memory errors are one-tenth
of the gate failure probabilities $\gamma_w=\gamma_{else}/10$,
we find a threshold value of $\gamma_{else}=3.4 \times 10^{-4}$.
This is smaller than what Steane estimates in Ref.\ \cite{steane:overhead}.
\item We find that, in the local setting,
the threshold scales as $\Theta(1/r)$. For example, for $r=20$ and
for the failure of movement over a unit distance equal to the
failure probability $\gamma_{else}$, and for memory errors equal to
one-tenth of $\gamma_{else}$, we find that the threshold value for
$\gamma_{else}$ is $7.3 \times 10^{-5}$.
\item We find that the threshold does not depend very strongly on the noise levels during
transportation.
\item We find that infrequent error correction may
have some benefits while qubits are in the `transportation
channel'.
\end{itemize}


\section{A Local Architecture}

Let us first discuss the existence of a fault-tolerance threshold
in the local model of quantum computation. It is clear that for
unencoded computations
 an at most a linear (in the number of qubits) overhead is incurred in order to make gates act on nearest-neighbor qubits.

\begin{figure}[htb]
\begin{center}
\epsfxsize=13cm \epsffile{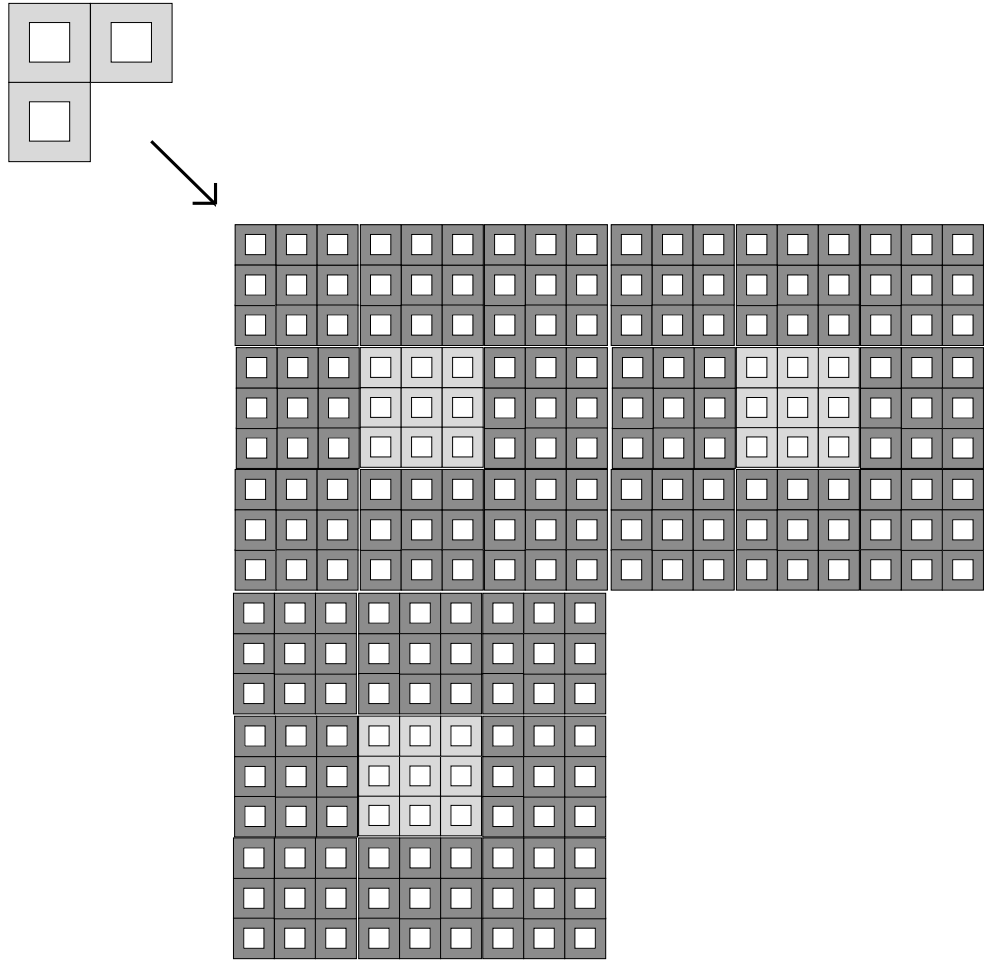} \caption{Two-dimensional plane with the spatial layout of $M_1$ and $M_2$. The grayness of the area indicates the amount of moving the qubits potentially need to do.}
\label{figschem}
\end{center}
\end{figure}

 If we perform concatenated coding in order to decrease the logical failure rate,
 we note that the circuit grows in size exponentially in the level of concatenation. Therefore,
 the distances over which qubits have to be transported (see
\footnote{It is understood here and elsewhere that transportation
of qubits does not necessarily mean transportation of the physical
embodiment of the qubit, but
 transport of the logical state of the qubit as in quantum teleportation.})
 and thus the number of places in time and space where errors can occur will increase. This will inevitably increase the
 logical failure rate at the next level of concatenation as compared to the logical failure rate in the nonlocal
 model. In order to be below the noise threshold, we want the logical failure rate to decrease at
 higher
 levels of concatenation. Thus it becomes a question of whether the extra increase in logical failure rate due
 to locality is sufficiently bounded so that there is still a noise value below which the logical failure rate
 decreases at the next level of concatenation. The question has been answered positively in the literature, see \cite{AB:faulttol,gottesman:localft}.
In particular, in Ref.\ \cite{gottesman:localft}, two simple, significant observations were made which are important
 in deriving the existence of a threshold in local fault-tolerant computation:
\begin{enumerate}
\item The most frequent operations during the computation should be the most local operations. For concatenated
 computation, the most frequent operation is lowest-level error correction. Thus the ancillas needed for this
 error correction should be adjacent to the qubits that are being corrected. The next most frequent is
 level 1 error correction, and so on. In Fig.\ \ref{figschem}, an example of a layout following these guidelines
 is given (see also \cite{gottesman:localft} itself).
\item The circuitry that replaces the nonlocal circuitry, say an
error correction routine or an encoded gate operation, should be
made
 according to the rules of fault-tolerance. For example, it is undesirable to swap a data qubit with another
 data qubit in the same block, since a failure in the swap gate will immediately produce two data errors. Local swapping could
 potentially be done with dummy qubits, whose state is irrelevant for the computation.
\end{enumerate}

The third observation, which is less explicitly stated in Ref.\ \cite{gottesman:localft}, is based
on the following. Let us assume that we follow the requirement for hierarchically putting error correction ancillas
near the data. We first start by making the original circuit a circuit with only nearest-neighbor gates
according to the specific architecture. We call this circuit $M_0$ and concatenate once to obtain circuit $M_1$, twice to
obtain circuit $M_2$, etc.
In circuit $M_1$, we have replaced qubits from $M_0$ by encoded qubits and their ancilla qubits for error correction
(or local gate operations). Thus every qubit becomes a `blob' of qubits with a certain spatial size. In order to
do a two-qubit gate $g$ from $M_0$, we have to move the data qubits in this blob past or over
these ancillary qubits in order to interact with other data qubits (see \footnote{Of course, for a single level of
concatenation, if $M_0$ is two-dimensional, we could put the ancillary qubits in the plane orthogonal to
the two-dimensional data plane, so that transversal two-qubit gates are between nearest-neighbor data blocks.
If we concatenate further however, we run out of dimensions.}).
Let us say that the scale of the blob is given by a parameter $r$ so that in order to do the encoded two-qubit
gate the qubits have to be moved over a distance $r$. At the next level of concatenation, again every
 qubit `point' becomes a blob, which implies that in order to do the doubly encoded version of $g \in M_0$,
 a doubly encoded block has to move over distance $r^2$. The two-qubit gates in the error correction
 of $M_1$ involving the level 1 error-correcting ancillas have to be moved over a distance $r$ and the
 level 0 error-correcting ancillas, which are added in $M_2$, are `local', assuming that we made the error correction
 routine itself local. Thus in general, in $M_n$, level $k$ ancillas, $k=0,\ldots, n-1$, may have to
 be moved over a distance which scales as $r^k$, exponential in the number of levels of concatenation.

 Let us assume that the failure probability of a travelling qubit is approximately
 linear in distance $d$, i.e.\ $p_{err}= 1-(1-p)^d \approx d p$ where $p$ is the failure probability per unit distance.
 For many implementations, the distances
 involved in moving level $k$ ancillas, as well as the failure rates, will be far too large and error correction
 will have to be done frequently while the qubits are in transit. In fact,
 a threshold will probably not even exist if
 there is no error correction done in transit. This is because at some level of concatenation
 the failure rates for the high-level ancillas are such that these ancillas
 completely decohere in transit. At that point, any additional level of
 concatenation can only make things worse, not better. In Section \ref{ftlocal:analyt}, we give the details of
 a model where (lower-level) error correction on `moving qubits' is included in the concatenation steps.

If we think about realistic architectures for any type of physical implementation, it is likely that the stationary
 qubits lie in a one-dimensional, two-dimensional, or a few stacks of two-dimensional planes, potentially
 clustered in smaller groups. The reason is that one likely needs
 the third dimension for the classical controls that act on the qubits as in ordinary computation.

 Given the discussion above, we can imagine a two-dimensional layout of qubits as in Fig.\ \ref{figschem}.
 In $M_1$, every block of data qubits surrounds stationary level 0 ancillas, indicated by the
 white area. The data qubits themselves have to be moved (over distance $r$) either out of the plane, or by `wires'
 in the plane, in order to interact with the nearest-neighbor block of data qubits.
 In $M_2$, we again have the stationary `white' level 0 ancillas,
 light gray areas for level 1 ancillas
 that now have to be moved over distance $r$, and the dark gray areas for data qubits
 which potentially have to be moved over distance $r^2$.

In this paper, we do not go into details about the mechanisms behind qubit movement. Inside the error correction
 procedure, depending on the implementation, one may think about swapping qubits or creating short-ranged
 EPR pairs in order to teleport qubits. For the longer distances, one may create a grid
 of EPR pairs, using quantum repeater techniques \cite{DBCZ:repeat}, which is maintained by frequent
 entanglement distillation, or alternatively convert stationary qubits
 into more mobile forms of qubits (photons, spin-waves, etc.). In Section \ref{ftlocal:analyt}, we lay out a model for
 error correction `along the way', but we do not discuss how or where in space this additional
 error correction can take place. This could be the subject of future research.

\section{Local Fault-Tolerance: An Analytic Lower Bound}
\label{ftlocal:analyt}

We follow the derivation of fault-tolerant quantum computation as in Ref.\ \cite{AB:ftsiam}, which has also
 been used in \cite{TB:ft} to deal with more general error models such as non-Markovian noise.

We denote the original quantum circuit as $M_0$, consisting
 of $N$ {\it locations}.  Each location is denoted by a triple
 $(\{q_0,\ldots,q_i\},U,t)$, where the set of $q_j$, $1\leq j \leq 2$, are the
 qubits involved in the operation $U$ at time $t$.  $U$ is restricted to
 one- and two-qubit gates for simplicity and can be the
 identity operation.
 We fix a computation code $C$ which encodes one qubit in $m$ qubits.
 To achieve a fault-tolerant circuit, we concatenate this code recursively $n$
 times to create the circuit $M_n$ that simulates, to $n$ levels of
 concatenation, the original circuit $M_0$.

The main change that occurs when including locality constraints in the fault-tolerance derivation is that
 additional `move' operations and error correction needs to be added. Secondly, the error correction
 procedure needs to be made local. How the latter task is done and what overhead is required will very much
 depend on the code. We will not focus on this issue in this paper.

\begin{figure}[htbp]
\mbox{
\Qcircuit @C=1em @R=.5em {
    & & & & & \lstick{\ket{p_0}} & \multigate{6}{\mathcal E} & \multigate{13}{\bf U}& \qw \\
    & & & & & \lstick{\ket{p_1}} & \ghost{E}& \ghost{G} & \qw \\
    & & & & & \lstick{\ket{p_2}} & \ghost{E} & \ghost{G} & \qw \\
    & & & & & \lstick{\ket{p_3}} & \ghost{E} & \ghost{G} & \qw \\
    & & & & & \lstick{\ket{p_4}} & \ghost{E} & \ghost{G} & \qw \\
    & & & & & \lstick{\ket{p_5}} & \ghost{E} & \ghost{G} & \qw \\
    \lstick{\ket{p}} & \multigate{1}{U} & \qw &
    \push{\rightarrow\rule{.3em}{0em}} & & \lstick{\ket{p_6}} &
    \ghost{E} & \ghost{G} & \qw \\
        \lstick{\ket{q}} & \ghost{U} & \qw & & & \lstick{\ket{q_0}} &
    \multigate{6}{\mathcal E} &\ghost{G} & \qw \\
    & & & & & \lstick{\ket{q_1}} & \ghost{E} & \ghost{G}& \qw \\
    & & & & & \lstick{\ket{q_2}} & \ghost{E} & \ghost{G}& \qw \\
    & & & & & \lstick{\ket{q_3}} & \ghost{E} & \ghost{G}& \qw \\
    & & & & & \lstick{\ket{q_4}} & \ghost{E} & \ghost{G}& \qw \\
    & & & & & \lstick{\ket{q_5}} & \ghost{E} & \ghost{G}& \qw \\
    & & & & & \lstick{\ket{q_6}} & \ghost{E} & \ghost{G}& \qw \gategroup{1}{7}{14}{8}{.6em}{--}\\
    & & & & & & & &\\
    & \mbox{$M_{n-1}$} & & & & & & \mbox{$M_n$}& & & &
}
}
\caption{The replacement rule for a two-qubit gate $U$.  The dashed box represents a 1-rectangle.  ${\mathcal E}$ represents the error correction procedure. ${\bf U}$ represents the encoded, fault-tolerant implementation of $U$.}
\label{nonlocal_replace}
\end{figure}
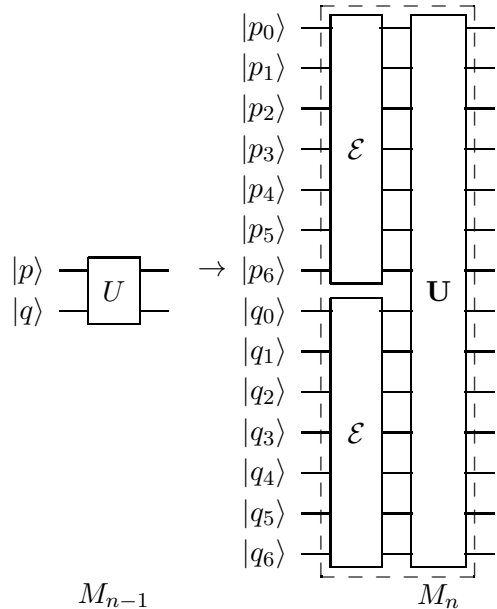

Consider a particular example of a location, for example a two-qubit gate.
 This gate gets replaced by a so-called 1-rectangle in $M_1$, which
 consists of error correction on both blocks of qubits followed by the encoded gate operation, shown in
 Fig.\ \ref{nonlocal_replace}.

\begin{figure}[htbp]
\mbox{ \Qcircuit @C=1em @R=.5em {
    & & & & & \lstick{\ket{p_0}} & \multigate{6}{\mathcal E} & \gate{{\rm
    move(r)}} & \multigate{6}{\mathcal E}& \multigate{13}{\bf U}& \multigate{6}{\mathcal E} & \gate{{\rm move(r)}} & \qw \\
    & & & & & \lstick{\ket{p_1}} & \ghost{E} & \gate{{\rm
    move(r)}} & \ghost{E}& \ghost{G} & \ghost{E} & \gate{{\rm move(r)}} & \qw \\
    & & & & & \lstick{\ket{p_2}} & \ghost{E} & \gate{{\rm
    move(r)}} & \ghost{E}& \ghost{G} & \ghost{E} & \gate{{\rm move(r)}} & \qw \\
    & & & & & \lstick{\ket{p_3}} & \ghost{E} & \gate{{\rm
    move(r)}} & \ghost{E}& \ghost{G} & \ghost{E} & \gate{{\rm move(r)}} & \qw \\
    & & & & & \lstick{\ket{p_4}} & \ghost{E} & \gate{{\rm
    move(r)}} & \ghost{E}& \ghost{G} & \ghost{E} & \gate{{\rm move(r)}} & \qw \\
    & & & & & \lstick{\ket{p_5}} & \ghost{E} & \gate{{\rm
    move(r)}} & \ghost{E}& \ghost{G} & \ghost{E} & \gate{{\rm move(r)}} & \qw \\
    \lstick{\ket{p}} & \multigate{1}{U} & \qw &
    \push{\rightarrow\rule{.3em}{0em}} & & \lstick{\ket{p_6}} &
    \ghost{E} & \gate{{\rm move(r)}} & \ghost{E}&
    \ghost{G} & \ghost{E} & \gate{{\rm move(r)}} & \qw
    \gategroup{1}{7}{7}{8}{.6em}{--}
    \gategroup{1}{11}{7}{12}{.6em}{--} \gategroup{1}{9}{14}{10}{.6em}{--} \\
        \lstick{\ket{q}} & \ghost{U} & \qw & & & \lstick{\ket{q_0}} &
    \multigate{6}{\mathcal E} & \gate{{\rm wait(r)}} &\multigate{6}{\mathcal E} &\ghost{G} & \multigate{6}{\mathcal E} & \gate{{\rm wait(r)}} & \qw \\
    & & & & & \lstick{\ket{q_1}} & \ghost{E} & \gate{{\rm wait(r)}} &\ghost{E}& \ghost{G}& \ghost{E} & \gate{{\rm wait(r)}} & \qw \\
    & & & & & \lstick{\ket{q_2}} & \ghost{E} & \gate{{\rm wait(r)}} &\ghost{E}& \ghost{G}& \ghost{E} & \gate{{\rm wait(r)}} & \qw \\
    & & & & & \lstick{\ket{q_3}} & \ghost{E} & \gate{{\rm wait(r)}} & \ghost{E}& \ghost{G}& \ghost{E} & \gate{{\rm wait(r)}} & \qw \\
    & & & & & \lstick{\ket{q_4}} & \ghost{E} & \gate{{\rm wait(r)}} & \ghost{E}& \ghost{G}& \ghost{E} & \gate{{\rm wait(r)}} & \qw \\
    & & & & & \lstick{\ket{q_5}} & \ghost{E} & \gate{{\rm wait(r)}} & \ghost{E}& \ghost{G}& \ghost{E} & \gate{{\rm wait(r)}} & \qw \\
    & & & & & \lstick{\ket{q_6}} & \ghost{E} & \gate{{\rm wait(r)}} & \ghost{E}& \ghost{G}& \ghost{E} & \gate{{\rm wait(r)}} & \qw \gategroup{8}{7}{14}{8}{.6em}{--} \gategroup{8}{11}{14}{12}{.6em}{--}\\
    & & & & & & & & & & & &\\
    & \mbox{$M_{n-1}$} & & & & & & & \mbox{$M_n$}& & & &
}} \caption{The replacement rule for a local two-qubit gate $U$. Each dashed box represents an elementary 1-rectangle.  ${\mathcal E}$ represents the error correction procedure. ${\bf U}$ represents the local fault-tolerant implementation of $U$.  The replacement circuit, i.e.\ the composite $1$-rectangle, contains five elementary 1-rectangles.} \label{2replace}
\end{figure}
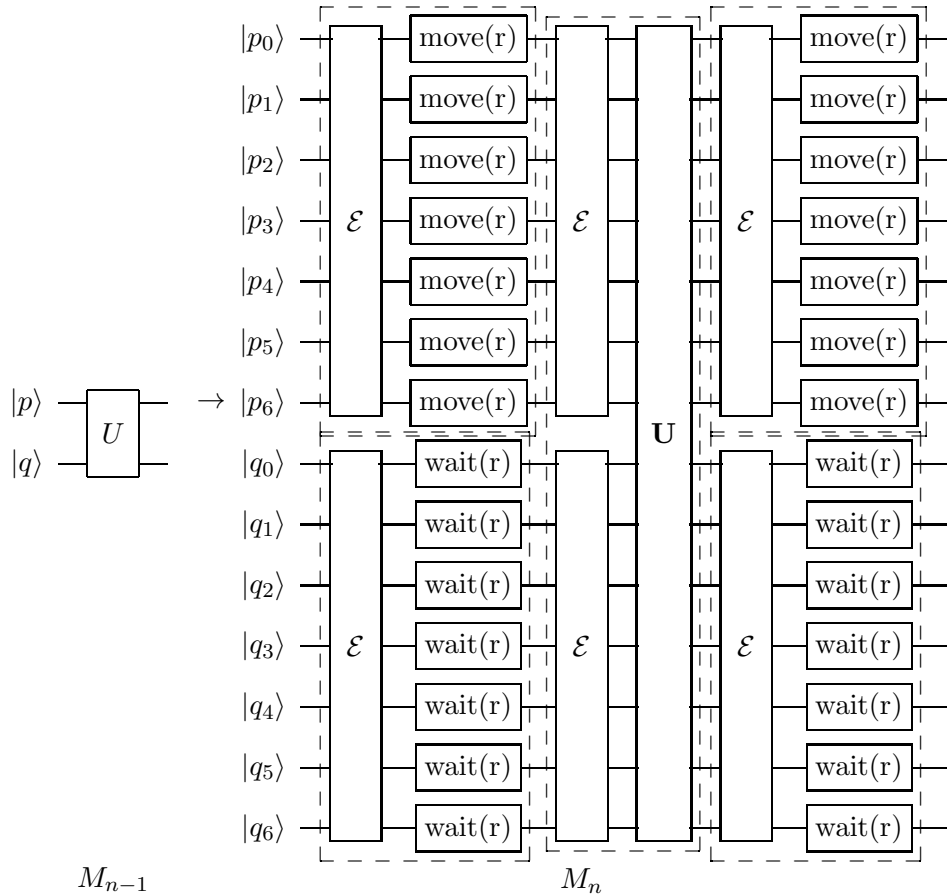

In the local model, this {\em replacement rule} that we repeatedly apply to obtain the circuit $M_n$
 gets modified as in Fig.\ \ref{2replace}. While one block gets moved over a distance $r$, which we denote as a ${\rm move(r)}$ operation,
 the other block is waiting. Next, the fault-tolerant implementation of the original gate is executed {\em locally} and then the block is moved back in place.
 We precede the move and wait operations by an error correction routine, just as for the gate $U$. The model that we consider here
 assumes that the error levels induced by moving over distance $r$ may be similar to the error levels
 due to the execution of the gate $U$. If moving is more error-prone, we may divide the distance $r$ into
 shorter segments of length $d$, $r= \tau d$, and error correct after every segment if necessary. This modification and its effects
 will be considered when we make our detailed analysis in Section \ref{local7}.

 We see that in the local model each location in $M_{n-1}$ gets replaced by potentially more than one `elementary'
 1-rectangle in $M_n$. Since this set of rectangles forms a logical unit, we will call the sequence
 of elementary 1-rectangles a {\em composite} 1-rectangle.

In the next section, we derive a rough lower bound on the
threshold in the local model,
 depending on a scale parameter $r$.


\subsection{Replacement Rules}\label{reprules}

We formulate replacement rules for all possible other locations in
the local model.  We only consider
 locations that occur in the $[[7,1,3]]$ code.
 Additional rules may have to be formulated for other codes,
 but the threshold estimate in this section will not depend on these details.  We assume in formulating these replacement rules
  that a one-qubit gate
 is never executed in parallel with a two-qubit gate
 (this is correct for the $[[7,1,3]]$ code that we study in Section \ref{local7}); this means that the execution of the one-qubit gate
 is not delayed by the additional moving required for the two-qubit gate.
 Note that we have two types of memory locations, which we call wait locations, depending on the type of gate (one- or two-qubit gate) occurring in the same time slice.
The figures depicting these rules can be found in Appendix A. Here
is a list of the distinct locations in the local model, also
listed in Table \ref{local_loc} in Section \ref{modchoices}, and
their replacement rules:
\begin{enumerate}
\item a one-qubit gate, depicted in Fig.\ \ref{wait1}.
\item a one-qubit gate followed by a measurement, depicted in Fig.\ \ref{1mreplace}. We group a measurement
 with a one-qubit gate, since the replacement rule for a measurement by itself is just doing $m$ measurements
 on $m$ encoded qubits.
\item a two-qubit gate $U$, depicted in Fig.\ \ref{2replace}.
\item a wait location in parallel with only one-qubit gates, denoted as ${\rm wait1}$ or $w1$. The replacement rule
 is the same as for a one-qubit gate (Fig.\ \ref{wait1}).
\item a wait location in parallel with two-qubit gates, denoted as ${\rm wait2}$ or $w2$, depicted in Fig.\ \ref{wait2}.
\item ${\rm move(r)}$, the operation which moves one qubit over distance
 $r$, where $r$ depends on code properties, depicted in Fig.\ \ref{move}.
\item ${\rm wait(r)}$, the operation which does nothing while another
 qubit moves over distance $r$, depicted in Fig.\ \ref{wait}.
\end{enumerate}

Note that our replacement rules enforce synchronization of gate
operations and waiting periods. Note that at each new level of
concatenation, every distance gets multiplied by the scale factor
$r$, so that a ${\rm move(r)}$ gate becomes $r$ ${\rm move(r)}$
1-rectangles. We would like to stress that the goal here has been
to choose a set of level-independent replacement rules that
capture the overall behavior; architecture, code-dependent and
concatenation-dependent optimizations are not considered.

In order to apply the rules repeatedly, the encoded gate $U$ is
broken down into local elementary gates (potentially using
additional swap gates) and the replacement rules are applied to
these local gates.

\subsection{Threshold Estimate}
\label{threshestim}

As was noted in Ref.\ \cite{AB:ftsiam} and explicitly stated in
Ref.\ \cite{TB:ft}, the formal derivation of a
 fault-tolerance threshold hinges on three Conditions (under the usual assumptions of having fresh ancillas
 and being able to operate gates in parallel). Fault-paths are subsets of all locations on which
 faults occur. The conditions are, loosely speaking, the
 following:
\begin{enumerate}
\item `Sparse' fault-paths (with few faults) lead only to sparse
errors. \item `Sparse' fault-paths give good final answers. \item
Non-sparse fault-paths have small probability/norm, going to zero
with increasing concatenation
 level for initial failure probabilities/norms per location below some threshold value.
\end{enumerate}

The first two statements are unchanged when going from a nonlocal to a purely local model of computation,
 assuming that the error correction routines are made local in a fault-tolerant manner. It is the
 third Condition whose
 derivation gets modified in this model. For concreteness, let us assume that our error model is a
 probabilistic error model, where each location undergoes a failure with some probability $\gamma(0)$.
 At an intuitive level, every location gets replaced by a composite 1-rectangle, which
 fails when at least one of the elementary 1-rectangles fails. If we assume that
 every type of 1-rectangle has a similar failure probability $\gamma(1)$, then the composite 1-rectangle which is
 most prone to failure is the one originating from the ${\rm move(r)}$ gate ($r >> 5$) since it consists of $r$
 elementary 1-rectangles. In order to be below the threshold,
 the failure probability of the composite 1-rectangle has to be
 smaller than the failure probability of the original location, i.e.
\beq
\gamma_0 \equiv \gamma(0) \geq 1-(1-\gamma(1))^{r} \approx \gamma(1) r.
\eeq
Let us assume that $A_{l,C}$ is (an upper bound on) the number of locations in an elementary 1-rectangle that has
 been made local. We say that a 1-rectangle fails if, say, more than $k$ among these locations have faults.
 Here $k=\lfloor d/2t\rfloor$ for a code with spread $t$ which can correct
 $d$ errors. Thus
 $\gamma(1) \approx {A_{l,C} \choose k+1} \gamma(0)^{k+1}$ and we get the threshold condition
\beq
{\gamma_0}_{\rm crit}=\frac{1}{\left(r {A_{l,C} \choose k+1}\right)^{1/k}}.
\label{thresval}
\eeq
The difference with the nonlocal model is the appearance of $r$ on the right-hand side of this equation. Note that the effects of locality
 {\em seem} to become effectively smaller for large $k$, i.e.\ for codes that can correct many errors. On the other hand, the scale factor $r$
 itself increases for codes that correct many errors, since the number of qubits in an encoded word and the size of the error-correcting machinery
 is larger. The $[[7,1,3]]$ code that we analyze in more detail
 in the next section does not entirely fit this analysis. The
 reason is that for the $[[7,1,3]]$ code, $d=1$ and $t=1$, causing $k$ to
 be zero; this is because one can have one incoming error (a late error
 in the previous rectangle) and one early error in the rectangle,
 leaving two errors on the data, which $[[7,1,3]]$ cannot
 correct. However a different analysis \cite{AGP:inprep} for such
 codes shows that one-error events in bigger `overlapping'
 rectangles (which include error-correction, gate operation and error-correction again)
 are acceptable for this code, so $k$ actually equals one.
 Thus we expect for the $[[7,1,3]]$ code that the threshold for the local
 model scales as $1/r$, which we partially confirm later.

A more formal analysis uses the notion of $n$-rectangles in $M_n$.
We state the definitions as given in Ref.\ \cite{AB:ftsiam} in
Appendix \ref{defft}.
In $M_n$, a composite $n$-rectangle originates from a single location in $M_0$. The composite $n$-rectangle
 consists of at most $r$ elementary $n$-rectangles. Each of these elementary $n$-rectangles consist of at most $A_{l,C}$
 composite $(n-1)$-rectangles, each of which again consists of at most $r$ elementary $(n-1)$-rectangles.
 Formally, we need to prove Condition (3) above, namely that the probability (assuming a probabilistic model)
 for sparse `good' faults gets arbitrarily close to one when we are below the threshold. Here we state the necessary
 lemma, which has identical structure to the one in \cite{AB:ftsiam}:

\begin{lemma}\label{smallerror}
If $\gamma_0 < {\gamma_0}_{\rm crit}, \exists \delta > 0$ such that the probability
 $\mathbb{P}(n)$ for the faults in a composite $n$-rectangle to be $(n,k)$-sparse is
 larger than $1-\gamma_0^{(1+\delta)^{n}}$.
\end{lemma}

\begin{proof}
Let $\delta$ be such that
\begin{equation}
r \displaystyle{A_{l,C} \choose {k+1}}\gamma_0^{k+1} < \gamma_0^{1+\delta}.
\label{belowthr}
\end{equation}
For $\gamma_0$ below the threshold, we can find such a $\delta$.  We prove
 the lemma by induction on $n$.  The probability for a composite 1-rectangle to have
 $(1,k)$-sparse faults, i.e.\ all elementary 1-rectangles (of which there are at most $r$)
 have sparse faults, is at least
\beq
\left(1-{A_{l,C} \choose k+1}\gamma_0^{k+1}\right)^r
\geq 1-r {A_{l,C} \choose k+1}\gamma_0^{k+1} > 1 - \gamma_0^{1+\delta},
\eeq
using Eq.\ (\ref{belowthr}).
Assume the lemma holds true for $n$ and we prove for $n+1$.
 For the faults in a composite $(n+1)$-rectangle not to be
 $(n+1,k)$ sparse, there must at least be 1 elementary $(n+1)$-rectangle in which
 the faults are not $(n+1,k)$-sparse which implies that in that rectangle
 there are at least $k+1$ composite $n$-rectangles which are not $(n,k)$-sparse.  Thus,
\beq
\mathbb{P}(n+1) \geq \left(1-{A_{l,C} \choose k+1}(1-\mathbb{P}(n))^{k+1}\right)^r \geq 1-r {A_{l,C} \choose k+1}(1-\mathbb{P}(n))^{k+1}.
\eeq
Using the induction hypothesis and Eq.\ (\ref{belowthr}) then gives
\beq
\mathbb{P}(n+1) > 1-(\gamma_0)^{(1+\delta)^{n+1}},
\eeq
as desired.
\end{proof}

We note that a similar analysis could be performed for any other noise model which is derived with the method used in
 Ref.\ \cite{AB:ftsiam}, such as noise satisfying the exponential decay conditions or local non-Markovian
 noise \cite{TB:ft}. The proof of Condition (3) in these cases needs to be altered to take into account
 the dependence on $r$.

\section{Nonlocal Fault-Tolerance For The 7-qubit $[[7,1,3]]$ Code}
\label{nonlocal7}

In order to make a good comparison between using a concatenated
$[[7,1,3]]$ code in the local or nonlocal model, we perform
 a fault-tolerance analysis for the nonlocal model. In Ref.\ \cite{steane:overhead}, Steane performed such an analysis and we follow his analysis to a certain
 extent.  At the end of this section, we summarize our findings for the nonlocal model. The goal
 is to produce a threshold in the right ballpark, taking into account various (but not all) details of the error correction
 circuitry. The details of error correction \cite{preskill:faulttol,steane:overhead, steane:fast, steane:space} are depicted in Figs.\ \ref{QEC}--\ref{S} in Appendix \ref{ecor}
 and can be described as follows.
 Error correction of a 7-qubit block consists of $X$- and $Z$-error correction denoted as $\mathcal{X}$ and $\mathcal{Z}$. For both types of
 error correction, one prepares $n_{rep}$ ancillas, using the $\mathcal{G}$ network in Fig.\ \ref{G_parallel}. These ancillas are tested for $X$ errors using the $\mathcal{V}$ network in Fig.\ \ref{V_parallel}
 and discarded if they fail the test. The probability for passing this test is called $\alpha$. If they do pass the test, they can be used to
 collect the syndrome as in Fig.\ \ref{S}. If the first collected syndrome is zero, then no further syndromes are collected (the idea being that it is likely
 that there is no error on the data). The probability for a zero syndrome is called $\beta$. If the syndrome is nonzero, an additional $s-1$
 syndromes are collected. These $s$ syndromes are then compared and if there are $s'$ among them which agree, then error recovery, denoted by
 $\mathcal{R}$, is done according to this syndrome. If there are no $s'$ which agree, no error correction is done and in our model (see
 \cite{steane:overhead} for modifications) we do not use these error syndromes in any subsequent error correction.

Let us now consider the problem of determining the fault-tolerance threshold by semi-analytical means. At the base-level, we start with
 a vector of failure probabilities of the locations in our model which we call $\vec{\gamma}(0)$. In our case we have the
 following five
 kinds of locations $l$; a one-qubit gate ($l=1$) with failure probability $\gamma_1 \equiv \gamma_1(0)$, a two-qubit gate ($l=2$) with failure
 probability $\gamma_2$, a wait location ($l=w$) with failure probability $\gamma_w$, a one-qubit gate followed by measurement ($l=1m$), with
 failure probability $\gamma_{1m}$, and a $\ket{0}$ preparation location with failure probability $\gamma_p$. Table \ref{nonlocal_loc} lists these
 types of locations in the nonlocal model.

\begin{table}[h]
\begin{tabular}{|c|c|c|}
Location  &  Description & Failure Prob.\\ \hline\hline $1$ &
one-qubit gate  & $\gamma_1$ \\ \hline $2$ & two-qubit gate &
$\gamma_2$ \\ \hline $w$ & memory (wait) & $\gamma_w$ \\ \hline
$1m$ & one-qubit gate $+$ measurement & $\gamma_{1m}$ \\ \hline
$p$ & preparation & $\gamma_p$ \\ \hline
\end{tabular}
\caption{Types of locations and their failure probability symbols
in the nonlocal analysis.} \label{nonlocal_loc}
\end{table}

There are several ways in which one can do a fault-tolerance
analysis. The first method is to perform a Monte-Carlo simulation
(see, for example,
 \cite{steane:overhead,reichardt:ft,zalka:ft}) of a sequence of operations
 for some level of concatenation and deduce a failure or crash probability. The advantage of this method is that it takes into account
 incoming errors into rectangles and then it otherwise exactly mimics the failure probability in the real quantum computation. The disadvantage,
 in particular for large codes, is that it is hard to simulate high levels of concatenation, since the size of the classical computation
 scales exponentially with concatenation level. As we discuss in a moment, and demonstrate in our studies, simulating more than one
 level of concatenation is often needed to nail down the threshold.

The second method is a semi-analytical one, which we follow, to
obtain an approximate probability flow equation.
 Due to concatenation, each location is represented by a rectangle, which has some probability of failure, meaning that at the end of the rectangle
 there are more errors on the data than the code can correct. Thus after one level of concatenation, the probability vector $\vec{\gamma}(0)$
 is mapped onto $\vec{\gamma}(1)$, and we repeat this procedure. We say that
 the original vector $\vec{\gamma}(0)$ is below the threshold if
$\vec{\gamma}(n) \rightarrow 0$ for large enough $n$. The drawback of this kind
of analysis is that careful approximations need to be made in order to estimate the failure probability function
 of a rectangle, since a complete analysis may be too complicated. Furthermore, the analysis does not deal so well with incoming errors, since we look at one 1-rectangle at a time.
 The advantage is that it is easy to look at high levels of concatenation.

In the next section, we approximate the failure probability function $\gamma_l(n)={\bf F}_l(\vec{\gamma}(n-1))$
 for the different types of 1-rectangles. First, we describe the modelling assumptions we have chosen.

\subsection{Modelling Choices}
\label{modchoices}

\begin{itemize}
\item We assume that the time it takes to do a measurement is the same
 as the one-qubit gate time and that classical post-processing does not take any additional time.
\item We have chosen to call a one-qubit gate followed by a
measurement a single location. The reason is
 that there is no explicit concatenation step for measurement, since each measurement just gets replaced
 by seven measurements and classical post-processing to correct for errors. We choose to set the failure
 probability of a measurement $\gamma_m=\gamma_1$. Thus the failure probability for the location $1m$
 is approximated as $\gamma_{1m} \approx \gamma_1+\gamma_m$. As it turns out, there are no
 two-qubit gates followed by measurement in the $[[7,1,3]]$ error correction routines, and a wait or memory location
 of any length followed by a measurement is just measurement, since there is no reason to wait.
 \item A preparation of the state $\ket{0}$ is a preparation location with a
 preparation failure probability $\gamma_p$. For simplicity, we may set $\gamma_{p}=\gamma_{1}$. At the next level of concatenation,
 this location will be replaced by an encoding circuit. Preparing an encoded
 $\ket{0}$ can be done by first performing error correction on an arbitrary state which projects the state into
 the code space and then measuring the eigenvalue of the encoded $Z$ operator fault-tolerantly and correcting if
 this eigenvalue is $-1$. Even though the last procedure, done fault-tolerantly, will be more involved
 than the execution of a transversal one-qubit gate, we assume that the encoding/preparation rectangle
 is of the one-qubit gate type. In other words, we do not use a separate replacement rule for a preparation location.
\item We will typically work in the regime where $\gamma_w <
\gamma_{1,2}$, perhaps an order of magnitude smaller. \item We
assume (here and in the local model) that our quantum circuit
contains only controlled-$Z$ ($C^Z$), controlled-not ($C^X$), and
Hadamard gates ($H$). Note that these can all be executed
transversally. Of course, in order to make the computation
universal, one would also need, e.g., a Toffoli gate or $\pi/8$
rotation. We believe that the inclusion of the $\pi/8$ gates would
not alter the threshold in the local model very much. The reason
is that (1) error-correction does not need the $\pi/8$ gate and
thus $\pi/8$ gates are fairly rare, (2) the $\pi/8$ gate, as a
1-qubit gate, can be executed locally and (3) the failure
probability for the 1-rectangle of the $\pi/8$-gate is probably
similar or lower than that of the 2-qubit gate since it involves
only one data block (and some ancillary state). As it turns out,
already the inclusion of the two-qubit gates has a sizable effect
on the threshold estimate.
\end{itemize}

The error correction procedure as described in the previous section is not of fixed size; for example,
 it depends on the number of syndromes collected and whether or not we do a recovery operation. Here are some choices that we make
 which directly affect how we calculate the failure probability in the next section. These assumptions are not exactly the same as the ones
 made in Ref.\ \cite{steane:overhead}:

\begin{itemize}
\item The procedures for error correction are of course parallelized as much as possible to reduce errors due
 to waiting. As can be seen in the figures, the syndrome collection network $\mathcal{S}$ (Fig. \ref{S}) then takes 3 time steps,
 the network ${\mathcal G}$ (Fig.\ \ref{G_parallel}) has 5 time steps and ${\mathcal V}$ (Fig.\ \ref{V_parallel}) has 6 time steps.
 We assume that the 4 verification bits are prepared while the ${\mathcal G}$ routine takes place.
\item We choose $s$, the maximum number of syndromes collected, to be $s=3$ and $s'=2$.
\item In every round of the computation, we assume that a nonzero syndrome occurs somewhere, so that in order to keep the network synchronized,
 the other data blocks have to wait for the additional $s-1$ syndromes to be collected. We take these wait locations into account.
\item We assume that a sufficient number $n_{rep}$ of new ancillas is
 prepared in parallel {\em before} the beginning of each error correction routine. We set $n_{rep}=\lceil \frac{s}{\alpha} \rceil$,
 so that on average we have enough ancillas for error correction. We assume that the ancillas are prepared during the previous error correction
 procedure so that the data does not have to wait in order to be coupled to the ancillas. These assumptions are a bit too optimistic, since
 a nonlocal ancilla preparation and verification routine, see Figs.\ \ref{G_parallel} and \ref{V_parallel}, takes 11 time steps, while three syndrome
 collection routines, see Fig.\ \ref{S}, take 9 time steps in total (and this will be worse in the local version of these procedures since
 ancillas have to be `moved in place' to couple to the data).
\item We assume that the prepared ancillas for the last $s-1$ syndrome collections have to wait before the previous
 syndrome collections are done. This could potentially be avoided, but we may as well include
 some extra wait locations since other approximations may be too optimistic.
\item In principle, we may not have enough syndromes in agreement, so that no error correction is performed, and secondly we could have enough
 syndromes agreeing but the syndrome may be faulty so that we do a faulty recovery operation. The latter probability may be quite small
 since errors have to `conspire' to make a faulty but agreeing syndrome, so we will neglect this source of errors. If we choose not to do
 error correction, we may have more incoming errors in the next routine; we do model incoming errors to some extent in our estimation
 of $\alpha$ and $\beta$, but we will not consider this source of errors separately.
\item In the estimation of the failure probability we always assume that faults do not cancel each other.
\item We are working with the probabilistic error model where each gate or location can fail with a certain probability. For a location on
 a single qubit that fails with probability $\gamma$, we say that a $X$, $Y$ or $Z$ error occurs with probability $\gamma/3$. We will use this
 distinction between $X$, $Y$ and $Z$ errors in our estimation of $\alpha$ and $\beta$ in the next section.
\end{itemize}

\subsection{Failure Probability}\label{failsource}

For the $[[7,1,3]]$ code failure of a 1-rectangle means that two
or more errors occur on data qubits during the execution of the
operations in the 1-rectangle.
 This could happen when we have a single incoming error and, say, a syndrome collection gate, such as $C^Z$, introduces an additional error
 on the data and the ancilla. In estimating the failure probability, we do not take into account incoming errors since below
 the threshold the probability for incoming errors should typically be small. The circuits are designed such that if there are
 no incoming errors and a single fault occurs in the 1-rectangle, that fault will typically either not affect the data, or will be corrected.
Only if the fault occurs late in the routine, say in the encoded gate operation, will the fault be passed on to the next error correction routine.
 Thus we assume that two faults affecting the data are needed for failure. First, let us consider those 1-rectangles which involve a single
 data block, i.e.\ $l=1, 1m, p,w$. Let ${\bf F}_l[s_x,s_z](\vec{\gamma})$ be the failure probability for a rectangle of type $l$ when
 $s_x$ and $s_z$ syndromes in resp.\ $\mathcal{X}$ and $\mathcal{Z}$ are calculated. We can write
\beq \gamma_l(n)=\beta^2 {\bf F}_l[1,1](\vec{\gamma}(n-1))
+2\beta(1-\beta){\bf F}_l[s,1](\vec{\gamma}(n-1))+(1-\beta)^2 {\bf
F}_l[s,s](\vec{\gamma}(n-1)). \label{1fail} \eeq From now on, we
will omit the dependence on concatenation level, i.e.\ we express
${\bf F}_l$ in terms of $\gamma_j$.
 Let $\mathbb{P}(e^+ \in T,s_x,s_z)$ be the probability of $e$ or more faults on the data block due to source $T$ when $s_x$ and $s_z$
 syndromes are calculated. We may model
\beq
\mathbb{P}(1^+ \in T,s_x,s_z)=1-(1-\delta(T))^{N(T,s_x,s_z)},
\label{1plus}
\eeq
where $\delta(T)$ is the failure
 probability of the particular location (or event) in $T$ which causes the fault and $N(T,s_x,s_z)$ counts the number of places in $T$ where the
 fault can occur. Similarly, we have
\beq
\mathbb{P}(2^+ \in T,s_x,s_z)=1-(1-\delta(T))^{N(T,s_x,s_z)}-\delta(T)N(T,s_x,s_z)(1-\delta(T))^{N(T,s_x,s_z)-1}.
\eeq
In Table \ref{failsourcetab}, we describe the possible sources of faults on the data and their values for $\delta$ and $N$. For failure to occur,
 we can typically have one fault due to source $I$ and one due to source $J$ or two faults due to source $I$. In other words, we approximate
\beq {\bf F}_l[s_x,s_z] \approx \sum_{I > J}\mathbb{P}(1^+ \in
I,s_x,s_z)\mathbb{P}(1^+ \in J,s_x,s_z)+ \sum_{I} \mathbb{P}(2^+
\in I,s_x,s_z). \label{failrect} \eeq  Some of the parts of the
first term give somewhat of an overestimate, since a single fault
in, say, $\mathcal{X}$ and a single fault
 in $\mathcal{Z}$ does not necessarily lead to a failure. Also, note that we are overcounting some higher order fault-terms, but these should
 be small. Note that the $l$ dependence of the right-hand side of Eq.\ (\ref{failrect}) only appears in
 the terms that involve the faults due to encoded
 gate operations listed in Table \ref{failsourcetab}. Note that we do not distinguish between $X$,$Y$ or $Z$ errors in estimating the failure
 probability.


\begin{table}[h]
\begin{tabular}{|c|c|c|}
                       Source                                                     & $\delta$  &   $N$   \\ \hline\hline
{\small Propagation from a verified ancilla with $X$ error}
&   $\delta_{\rm anc}$  & $s_x+s_z$                 \\ \hline
{\small Fault in $C^Z$ or $C^X$ in $\mathcal{S}$}
&    $\gamma_2$      & $7(s_x+s_z)$       \\ \hline {\small Memory
faults on data at the end of $\mathcal{S}$}           & $\gamma_w$
& $14(s_x+s_z)$ \\ \hline {\small Memory faults on data during
$\mathcal{R}$}                  & $\gamma_w$ &
$6(\delta_{s_z,s}+\delta_{s_x,s})$ \\ \hline {\small Fault in gate
of $\mathcal{R}$}                              &
$\gamma_1+\gamma_{ws}$ & $\delta_{s_z,s}+\delta_{s_x,s}$ \\ \hline
{\small Memory faults on data when $s=1$}
& $\gamma_w$ & $21(s-1)(\delta_{s_z,1}+\delta_{s_x,1})$\\ \hline
{\small $X$ errors on ancillas waiting for $\mathcal{S}$}
& $\gamma_w$ & $21 s(s-1)(\delta_{s_z,s}+\delta_{s_x,s})/2$ \\
\hline
{\small Encoded gate error in rect. of type $l$}                        & $\gamma_l$ & $7$ \\
\hline
\end{tabular}
\caption{Different sources of failure and their contribution to the failure probability. Here $\delta_{\rm anc}=1-\mathbb{P}(\mbox{no $X$ }|\mbox{ pass})$ where $\mathbb{P}(\mbox{no $X$ }|\mbox{ pass})=\mathbb{P}(\mbox{pass and no $X$})/\alpha$ and $\mathbb{P}(\mbox{pass and no $X$})$ is the probability that an ancilla passed verification {\em and} has no $X$ errors on it. This probability is estimated in Section \protect\ref{estmab}. The probability $\gamma_{ws}$ for obtaining a wrong majority syndrome is assumed to be 0 in our analysis. }
\label{failsourcetab}
\end{table}

For a $l=2$ ($C^X$ or $C^Z$) 1-rectangle the analysis is slightly more involved. Let
 ${\bf F}[s_{x_1},s_{z_1},s_{x_2},s_{z_2}]$ be the failure probability of the two error correction routines
 on block 1 and 2 when $s_{x_1}$ and $s_{z_1}$ syndromes are computed for block 1 and
 $s_{x_2}$ and $s_{z_2}$ syndromes are computed for block 2 (without the subsequent gate operation).
Let $m_{j} \in \{0,1\}$ such that $m_j=0$ when $s_{j}=s$ and $m_j=1$ when $s_j=1$, where $j \in \{x_1,x_2,z_1,z_2\}$ and $s$ is the number of syndrome measurements.
We can then write
\bea
\gamma_2(n)=\sum_{s_{x_1},s_{z_1},s_{x_2},s_{z_2}=1,s}
(\beta)^{m_{x_1}+m_{x_2}+m_{z_1}+m_{z_2}}\times \nonumber \\
(1-\beta)^{4-m_{x_1}-m_{x_2}-m_{z_1}-m_{z_2}}
{\bf F}[s_{x_1},s_{z_1},s_{x_2},s_{z_2}](\vec{\gamma}(n-1)).
\eea
Let ${\bf F}(s_x,s_z)$ be the failure probability of one error correction routine when $s_x$ and $s_z$ syndromes
 are calculated, i.e.\ it is Eq.\ (\ref{failrect}) with the additional constraint that the source is never the encoded gate.
 Let $\mathbb{P}(1^+ \in T,s_{x_1},s_{z_1},s_{x_2},s_{z_2})$ be the probability of one or more faults anywhere due to source
 $T$ in the two error correction routines calculating $s_{x_1},s_{z_1},s_{x_2},s_{z_2}$ syndromes, that is, the number $N$ in Eq.\ (\ref{1plus})
 gets modified to $N(T,s_{x_1},s_{z_1},s_{x_2},s_{z_2})$ which is similar to the ones in Table \ref{failsourcetab} except that we add
 the contributions from both error corrections. Then for $C^A$, where $A=X$ or $A=Z$, we approximate
\bea
{\bf F}(s_{x_1},s_{z_1},s_{x_2},s_{z_2}) \approx \mathbb{P}(2^+ \in C^A)+7 \gamma_2 (1-\gamma_2)^6 \sum_{I \neq G}
\mathbb{P}(1^+ \in I,s_{x_1},s_{z_1},s_{x_2},s_{z_2})
+\nonumber \\ (1-\gamma_2)^7 [{\bf F}(s_{x_1},s_{z_1})+{\bf F}(s_{x_2},s_{z_2})].
\label{2fail}
\eea
The first term represents the contribution from having two or more faults in the two-qubit gate, the second term represents one gate fault
 and one or more faults somewhere in the error correction routines and the third term represents no faults in the gates
 and two or more in either the error correction on block 1 or block 2.


\subsection{Estimation of $\alpha$ and $\beta$}\label{estmab}

Our next task is to provide estimates for $\alpha$, the probability of an ancilla passing verification, and $\beta$, the probability of
obtaining a zero syndrome. One can
 find another estimation of $\alpha$ and $\beta$ in Ref.\ \cite{steane:overhead}. Similar to the failure probability, $\alpha$ and $\beta$ are functions
 of concatenation level, i.e.\ ${\bf F}_l(\vec{\gamma}(n-1))$ involves the functions $\alpha(n-1) \equiv \alpha(\vec{\gamma}(n-1))$
 and $\beta(n-1) \equiv \beta(\vec{\gamma}(n-1))$. In the following we omit the concatenation level dependence, i.e.\ we express
 $\alpha$ and $\beta$ in terms of $\gamma_i$.

For CSS codes, error correction is performed in two steps. While
$X$ and $Z$ errors are detected in only one of the two steps, $Y$
errors contribute to both. Hence if $X,Y,Z$ errors are equally
likely, the probability to detect an error is $2/3p$ for each
step, where $p$ denotes the total error probability.

In the following paragraph we will speak of events that are
detected as $X$ errors or $Z$ errors. Thus if a $Y$ error occurs
this results in both an $X$ and $Z$ error event.


The fraction $\alpha$ of ancillas that pass verification can be calculated as
\bea
\alpha=\mathbb{P}(\mbox{pass and no $X$})+\mathbb{P}(\mbox{pass and $X$})= \nonumber \\
\mathbb{P}(\mbox{pass and no $X$})+\mathbb{P}(\mbox{pass and no $Z$})- \nonumber \\
\mathbb{P}(\mbox{pass and no $Z$, no $X$})+\mathbb{P}(\mbox{pass and $Z$,$X$}).
\eea
The last probability we approximate as $\mathbb{P}(\mbox{pass and $Z$,$X$}) \approx 0$. The next table shows
 what types of errors should be avoided in order to have a passing ancilla and no $X$ or no $Z$ errors.


\begin{table}[h]
\begin{tabular}{|c|c|c|c|c|c|c}
                                   & {\small prep. ver. bits}  &   {\small $H$+meas. ver. bits}  & {\small from $\mathcal{G}$} &  {\small early wait in $\mathcal{V}$} &  {\small late wait in $\mathcal{V}$} & {\small ver. wait in $\mathcal{V}$} \\ \hline\hline
{\small $\mathbb{P}(\mbox{pass and no $Z$})$} &       $X$,$Z$ &
$X$                &         $X$,$Z$          &$X,Z$   & $Z$   &
$X$,$Z$
\\ \hline
{\small $\mathbb{P}(\mbox{pass and no $X$})$} &         $Z$        &           $X$               & $X$ &         $X$                & $X$ & $Z$               \\
\hline
\end{tabular}
\caption{Types of errors in various subroutines that should not
occur when ancilla passes verification and should have no $X$ or
no $Z$ errors. When we write $Z$, it implies that neither $Z$ nor
$Y$ should occur, since $Y$ is both an $X$ and $Z$ error. Late
wait indicates the wait locations on ancilla qubits that are
finished interacting with the verification qubits. Early wait
locations indicate the wait locations that occur before the last
interaction with the verification bits. Verification wait
locations indicate the wait locations that occur on the
verification qubits. Strictly speaking, for the contribution to
$\mathbb{P}(\mbox{pass and no $Z$})$ we should distinguish between
early and late wait errors on the verification qubits; we
approximate this by requiring no types of errors on the
verification qubit wait locations.} \label{errorcont1}
\end{table}


For the $C^Z$ gates, the exact contributions from various errors is harder to estimate (one has to examine the
 cases more carefully), so we approximate this by saying that in order to have a passed ancilla
 and no $Z$ or no $X$ error on the ancilla, all $C^Z$ gates have to have no errors.
 This implies that
\bea
\mathbb{P}(\mbox{pass and no $Z$})=(1-\gamma_p)^4 (1-\gamma_1)^4 (1-2\gamma_{1m}/3)^4 \times \nonumber \\
\Pi_{i \in \mathcal{G}}(1-\gamma_i)^{N(i \in \mathcal{G})}
(1-2\gamma_w/3)^{26} (1-\gamma_w)^6 (1-\gamma_2)^{13},
\label{zanc} \eea

and, slightly different,

\bea
\mathbb{P}(\mbox{pass and no $X$})=(1-2\gamma_p/3)^4 (1-2\gamma_1/3)^4 (1-2\gamma_{1m}/3)^4 \times \nonumber \\
\Pi_{i \in \mathcal{G}}(1-2\gamma_i/3)^{N(i \in
\mathcal{G})}(1-2\gamma_w/3)^{32} (1-\gamma_2)^{13}. \label{xanc}
\eea

Assuming that none of the possible faults occurs, then we can say
that \bea \mathbb{P}(\mbox{pass and no $Z$, no $X$}) \approx
\Pi_i(1-\gamma_i)^{N(i \in \mathcal{G,V})}. \label{noxz} \eea From
these estimates we can calculate $\alpha$.

Next we approximate $\beta$, the probability of obtaining a zero syndrome, in a $X$-error correction routine as
\bea
\beta \approx \mathbb{P}(\mbox{no $Z$ errors on anc. }|\mbox{ ancilla passed})\times
\mathbb{P}(\mbox{no $Z$ errors on syn. due to ${\mathcal S}$})\times \nonumber \\
\mathbb{P}(\mbox{no $X$ error coming into }\mathcal{X}).
\label{eqn:beta1}
\eea
We have
\beq
\mathbb{P}(\mbox{no $Z$ errors on anc. }|\mbox{ ancilla passed})=\mathbb{P}(\mbox{pass and no $Z$})/\alpha.
\eeq
It is easy to estimate
\beq
\mathbb{P}(\mbox{no $Z$ errors on syn. due to ${\mathcal S}$})=(1-2\gamma_2/3)^7 (1-2\gamma_{1m}/3)^7.
\eeq
Thirdly, we have
\bea
\mathbb{P}(\mbox{no incoming $X$ error in }\mathcal{X})=
\mathbb{P}(\mbox{no incoming $X$ error in }\mathcal{Z})\times \nonumber \\
\left[\beta\mathbb{P}(\mathcal{S}_1 \in \mathcal{Z} \mbox{ leaves no $X$ error})
\mathbb{P}(\mbox{no $X$ err. on waiting data}) \right.+\nonumber \\
\left.(1-\beta)\mathbb{P}(\mathcal{S}_{1,2, \ldots,s} \in \mathcal{Z}
\mbox{ leave no $X$ error})\right], \eea

What is the probability $\mathbb{P}(\mbox{no incoming $X$ errors in } \mathcal{Z})$? If we assume that
 the previous $\mathcal{X}$ did its job, i.e.\ removed the errors, the only source of error is the gate
 that was done after $\mathcal{X}$. Since we do not know which gate was performed, we assume that
 the most error-prone gate occurred. Since all gates in our model are transversal, we approximate
\beq
\mathbb{P}(\mbox{no incoming $X$ errors in } \mathcal{Z}) \approx (1-2(\max_i \gamma_i)/3)^7.
\label{ince}
\eeq
We further estimate
\bea
\mathbb{P}(\mathcal{S}_1 \in \mathcal{Z} \mbox{ leaves no $X$ error}) =
\mathbb{P}(\mathcal{S}_1 \mbox{ gives no $X$ errors on data})\times \nonumber \\
 \mathbb{P}(\mbox{no $X$ errors on anc. }|\mbox{ anc. passed}).
\eea
where
\beq
\mathbb{P}(\mathcal{S}_1 \mbox{ gives no $X$ errors on data})=(1-2\gamma_2/3)^7.
\eeq

Lastly, we have
\beq
\mathbb{P}(\mbox{no $X$ errors on anc. }|\mbox{ anc. passed})=\mathbb{P}(\mbox{pass and no $X$})/\alpha.
\eeq
We also estimate
\bea
\mathbb{P}(\mathcal{S}_{1,2, \ldots,s} \in \mathcal{Z} \mbox{
leave no $X$ error}) \approx \mathbb{P}(\mathcal{S}_{1} \in
\mathcal{Z} \mbox{ leave no $X$ error})^s.
\eea
This estimate does not include the fact that the prepared ancillas may have to wait
 (and degrade) until they are coupled to the data. If there is only one syndrome collection, the data may have to
 wait until other full syndrome collections are done. We take this into account with
\beq \mathbb{P}(\mbox{no $X$ err. on waiting
data})=(1-2\gamma_w/3)^{21(s-1)}. \label{waitdat} \eeq Thus, using
Eqs. (\ref{eqn:beta1}) -- (\ref{waitdat}), we arrive at a closed
formula for $\beta$.

\section{Numerical Threshold Studies for The Nonlocal Model}

We have used the formulas for failure probabilities, $\alpha$, and
$\beta$ of the last two subsections to quantify the
fault-tolerance threshold for the nonlocal model. We study the
effect of the repeated application of the map ${\bf
F}_l(\vec{\gamma})$, namely the dependence of the parameters on
concatenation level. This is a four-dimensional map --- there are
five probability variables, but under our assumptions $\gamma_1$
and $\gamma_p$ behave identically.
This four-dimensional flow is of course impossible to visualize
directly, but two-dimensional projections of these flows prove to
be very informative.

\begin{figure}[htb]
\begin{center}
\epsfxsize=15cm \epsffile{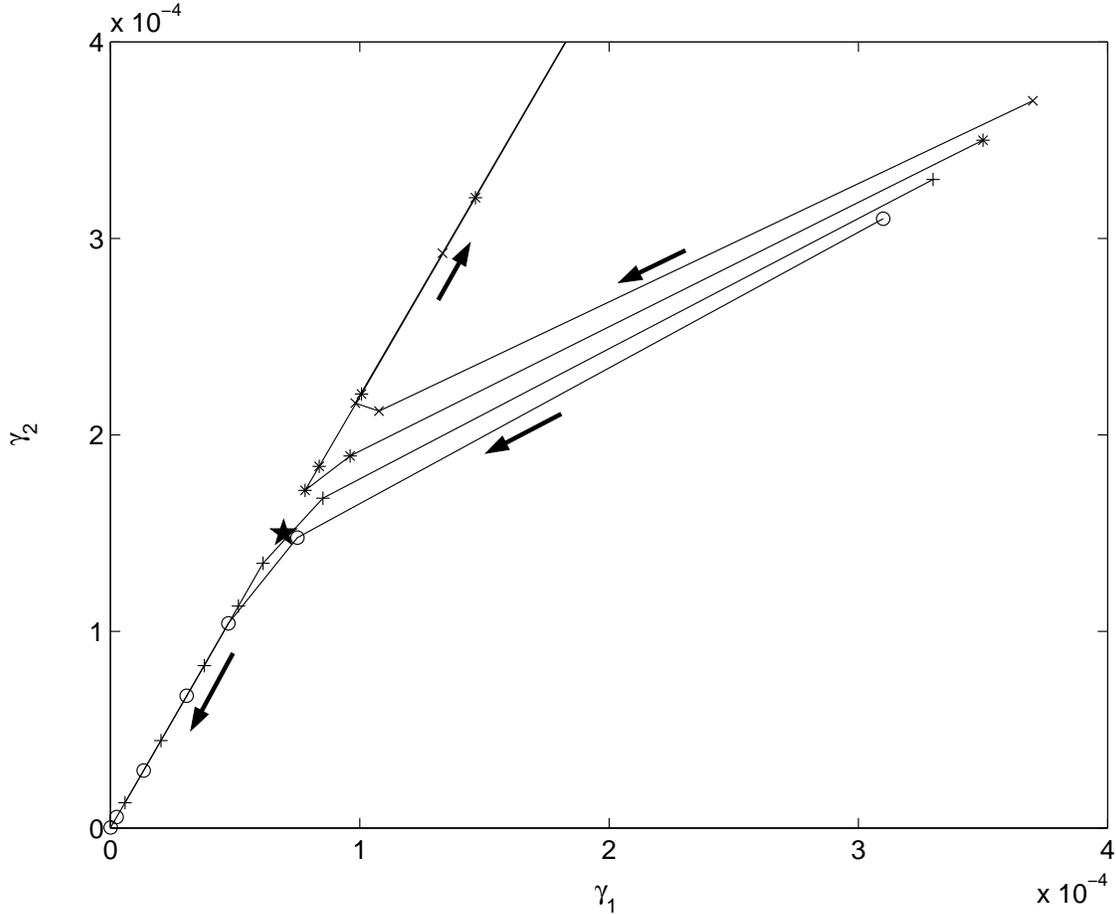}
\caption{Flows of the one- and two-qubit failure rates under concatenation in the nonlocal model. We initially set $\gamma_1=\gamma_2=\gamma_p=\gamma_m=10\times\gamma_w$. Four starting values are shown, two below threshold and two above.
The initial flow is evidently very similar regardless of whether the map is above or below threshold.  The hyperbolic structure of the flow is controlled by an unstable fixed point of the map at $\gamma_1=\gamma_w=\gamma_{1m}=\gamma_p=0.69\times 10^{-4}$, and $\gamma_2=1.50\times 10^{-4}$, shown as the black ``star" symbol. Note that the line onto which these flows asymptote has $\gamma_2$ very close to $2\times\gamma_1$.}
\label{XXX}
\end{center}
\end{figure}

\begin{figure}[htb]
\begin{center}
\epsfxsize=15cm \epsffile{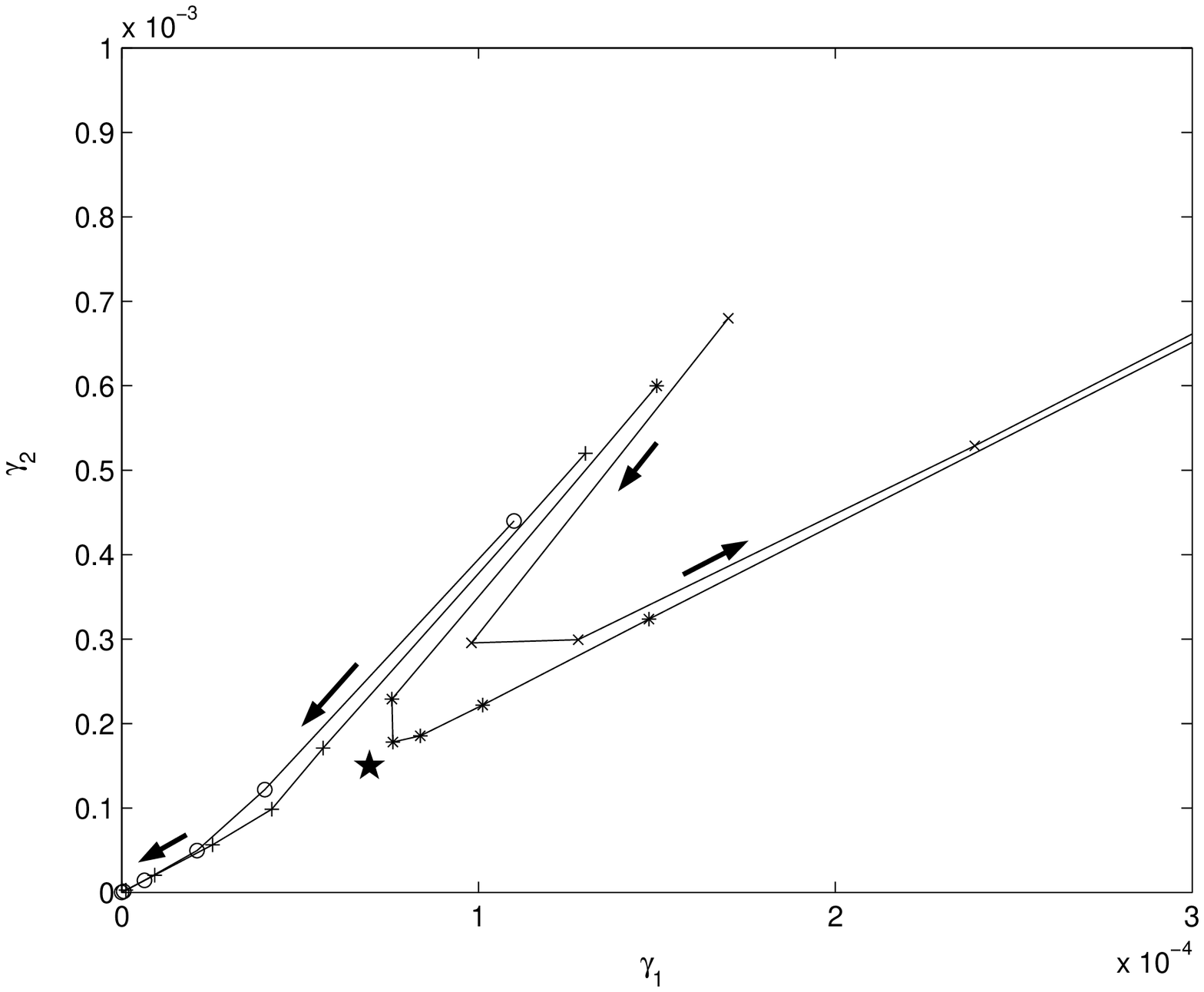}
\caption{
Flows of the one- and two-qubit failure rates under concatenation in the nonlocal model.
We initially set $\gamma_1= 0.25 \times \gamma_2=\gamma_p=\gamma_m=10\times\gamma_w$.
Four starting values are shown, two below threshold and two above. The initial flow is evidently very similar regardless of whether the map is above or below threshold.  The hyperbolic structure of the flow is controlled by an unstable fixed point of the map at $\gamma_1=\gamma_w=\gamma_{1m}=\gamma_p=0.69\times 10^{-4}$, and $\gamma_2=1.50\times 10^{-4}$, shown as the black ``star" symbol. Note that the line onto which these flows asymptote has $\gamma_2$ very close to $2\times\gamma_1$.}
\label{quarter}
\end{center}
\end{figure}

\begin{figure}[htb]
\begin{center}
\epsfxsize=15cm \epsffile{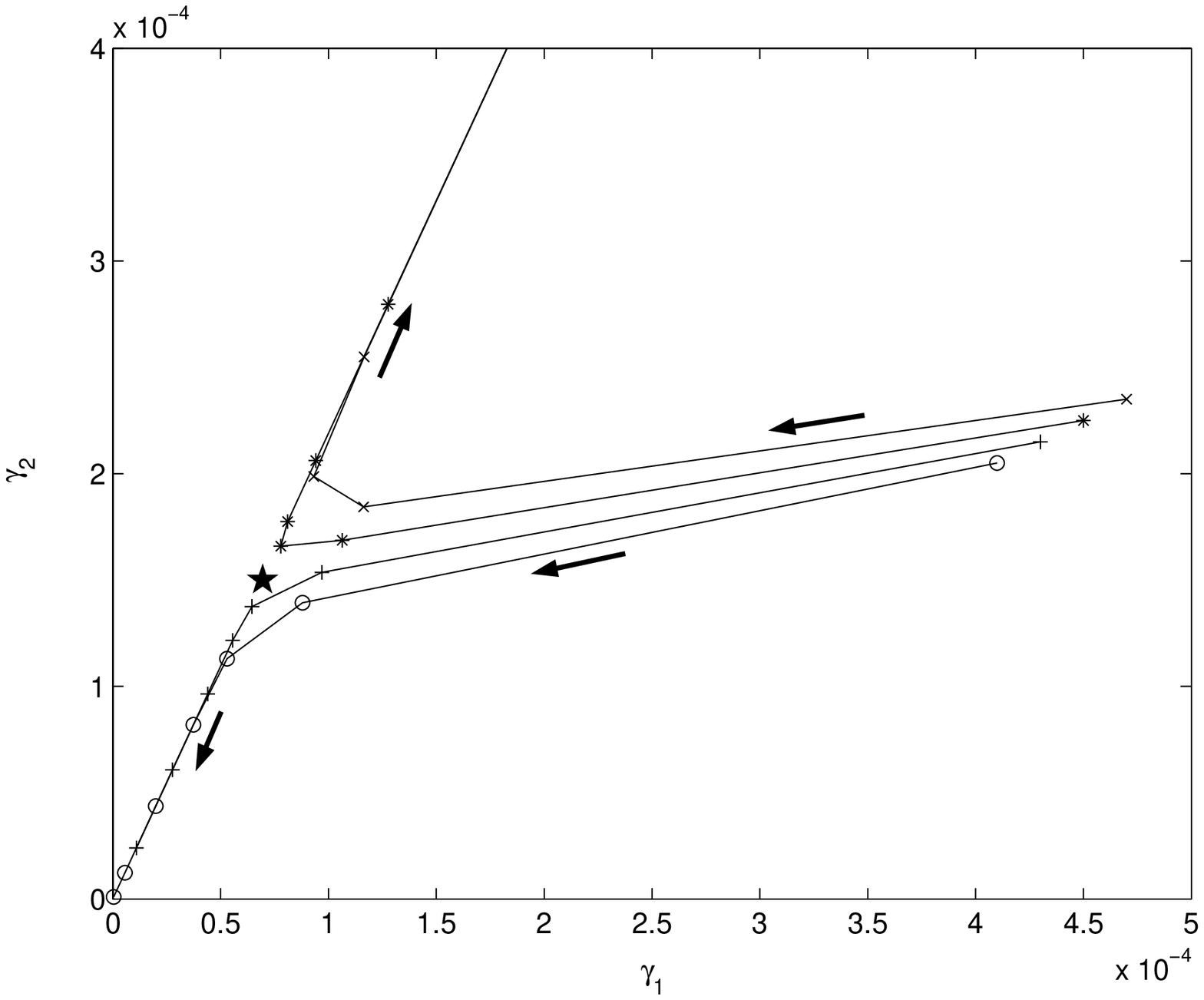}
\caption{
Flows of the one- and two-qubit failure rates under concatenation in the nonlocal model.
We initially set $\gamma_1= 2.0 \times \gamma_2=\gamma_p=\gamma_m=10\times\gamma_w$.
Four starting values are shown, two below threshold and two above. The initial flow is evidently very similar regardless of whether the map is above or below threshold.  The hyperbolic structure of the flow is controlled by an unstable fixed point of the map at $\gamma_1=\gamma_w=\gamma_{1m}=\gamma_p=0.69\times 10^{-4}$, and $\gamma_2=1.50\times 10^{-4}$, shown as the black ``star" symbol. Note that the line onto which these flows asymptote has $\gamma_2$ very close to $2\times\gamma_1$.}
\label{twice}
\end{center}
\end{figure}

In Figures \ref{XXX} -- \ref{twice} we show three instances of
such a projected flow in the $\gamma_1-\gamma_2$ plane.  In Fig.\
\ref{XXX} we have initially taken the memory failure probability
to be 10\% of the gate failure probability and one- and two-qubit
gate failure probabilities to be equal; that is, prior to
concatenation, we take
$\gamma_1=\gamma_2=\gamma_p=\gamma_m=10\times\gamma_w$. In Fig.\
\ref{quarter}, we initially take $\gamma_1=0.25 \times
\gamma_2=\gamma_p=\gamma_m=10\times\gamma_w$.  In Fig.\
\ref{twice}, we initially take $\gamma_1=2.0 \times
\gamma_2=\gamma_p=\gamma_m=10\times\gamma_w$. With these initial
choices, we look at the flows as we concatenate the map. Figures
\ref{XXX} -- \ref{twice} show the behavior as the threshold noise
value is crossed. As is common in renormalization group flows,
these have a hyperbolic character; the flows all asymptote to a
one-dimensional line (for which, as can be seen in the figures,
$\gamma_2\approx 2\gamma_1$). In Fig.\ \ref{XXX}, for all initial
points up to $\gamma_2\leq 3.35\times 10^{-4}$, the flows follow
this line to the origin, indicating successful fault-tolerant
computation; for all higher failure rates the flows asymptote to
one, indicating the failure of error correction.

The whole character of the flow is set by the presence of an
unstable fixed point at the black star, at approximately $\gamma_1=\gamma_w=\gamma_{1m}=\gamma_p=0.69\times
10^{-4}$, and $\gamma_2=1.50\times 10^{-4}$ in Figs.\ \ref{XXX} -- \ref{twice}.  It is evident that
the linearized map around this point has one positive (unstable)
eigenvalue and four negative ones.

\begin{figure}[htb]
\begin{center}
\epsfxsize=15cm \epsffile{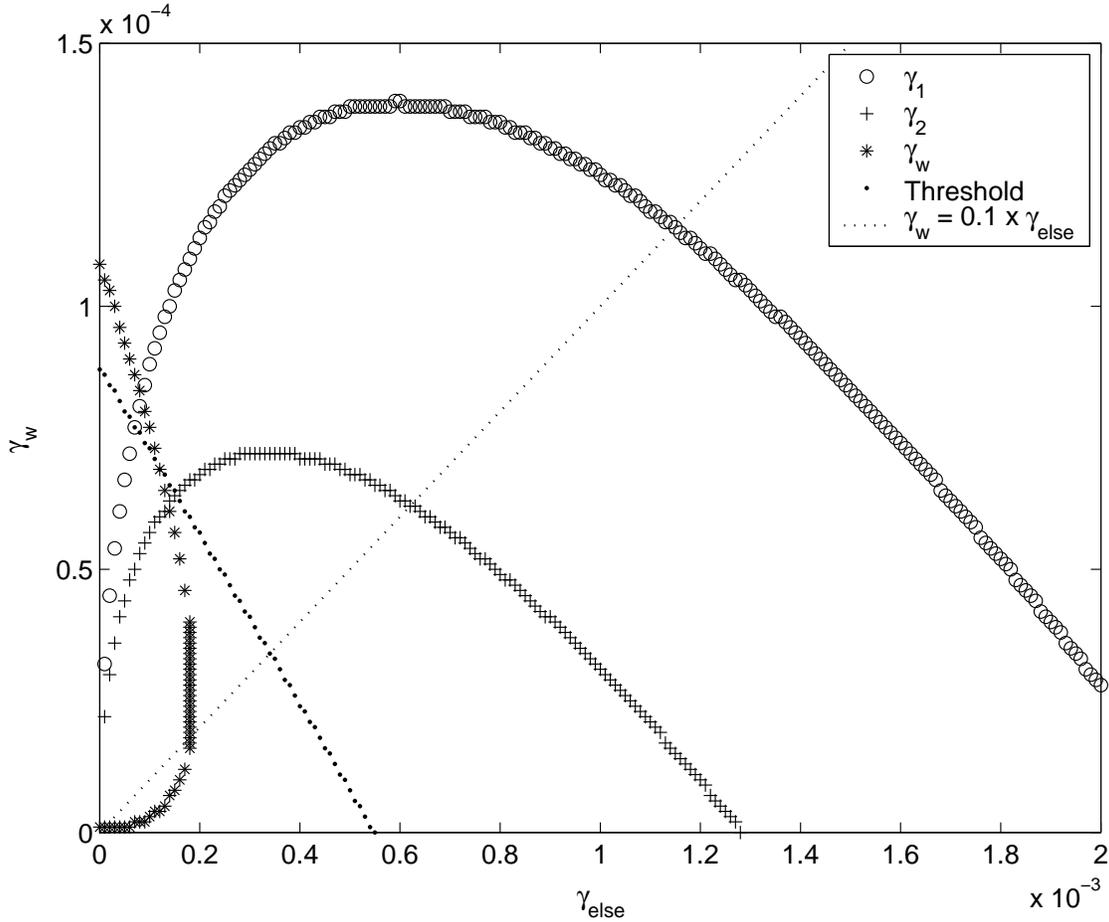}
\caption{The threshold line and pseudothreshold curves shown in the plane defined by the memory failure rate $\gamma_w$ and all other failure rates $\gamma_{else}=\gamma_1=\gamma_2=\gamma_p=\gamma_m$.  The pseudothreshold is defined as the line along which one of the failure rates remains unchanged after the first iteration of the map; closer to the origin, this failure rate decreases, further away it increases.  The pseudothresholds for $\gamma_1$, $\gamma_2$, and $\gamma_w$ are shown.  We note that along the line (dotted) for which $\gamma_w=0.1\times\gamma_{else}$, a popular condition in earlier studies, the gate pseudothresholds, particularly for the one-qubit gate failure rate, are much higher than the true threshold.}
\label{YYY}
\end{center}
\end{figure}

The threshold, of course, is not a single number; it is the
separatrix between points in the four-dimensional space of failure
probabilities that flow to the origin upon concatenation, and
those that flow to one. This separatrix is a three-dimensional
hypersurface. A one-dimensional cut through this hypersurface is
shown in Fig.\ \ref{YYY}. This is shown in the plane of memory
failure $\gamma_w$ versus all other failures, with all these rates
taken to be the same:
$\gamma_{else}=\gamma_1=\gamma_2=\gamma_p=\gamma_m$. The threshold
curve (indicated with black 'dot' symbols) is nearly approximated
by a straight line.

In Ref.\ \cite{steane:overhead} it has been suggested that a
reasonable estimate for the threshold can be obtained by finding
the failure rate for which the error is unchanged after the first
concatenation of the error-correcting code.  Figures \ref{XXX} and
\ref{YYY} indicate that this rule of thumb actually has limited
value (see \footnote{If all failure probabilities were the same, and stayed the same
after concatenation (this is what we assume in the analytical estimate
in Section \ref{ftlocal:analyt}), it is correct that one level of concatenation
is enough to obtain the threshold.}).
For all plotted initial points in Fig. \ref{XXX}, the failure
probabilities go down after one level of concatenation. However,
after one more level of concatenation, two of the failure
probabilities go up again
 indicating that those two initial points were above threshold.

In Fig.\ \ref{YYY} we investigate this further by plotting three
``pseudothresholds" along with the actual threshold curve. These
pseudothresholds are the lines along which $\gamma_1$, $\gamma_2$,
and $\gamma_w$ are unchanged after one iteration of the map.
Obviously, these three are very different from one another and
from the true threshold curve.  Rather than being straight, the
pseudothresholds are very curved.  They curve in to the origin for
a very simple reason: if $\gamma_1$ and $\gamma_2$ are initially
zero, then no matter what the value of $\gamma_w$ (i.e.\ anywhere
along the $y$-axis of the plot), $\gamma_1$ and $\gamma_2$ become
nonzero after one iteration, so every point on the $y$-axis is
above these pseudothresholds.  The corresponding statements hold
about the $x$-axis for the $\gamma_w$ pseudothreshold.

We note that, particularly in the region where $\gamma_w <<
\gamma_{else}$, the $\gamma_1$ pseudothreshold is a very
substantial overestimate of the true threshold.  On the plot we
indicate the line for which memory failure is one-tenth of gate
failure, a situation studied extensively by Steane
\cite{steane:overhead}. The $\gamma_1$ pseudothreshold is around
$\gamma_{else}=1.2\times 10^{-3}$ (near the threshold value
estimated by Steane), while the true threshold is at
$\gamma_{else}=0.34\times 10^{-3}$, about a factor of four lower.
Looking at a wider range of initial failure rate values, we find
that the initial point $\vec{\gamma}(0)$ is below its true
threshold whenever all of the $\gamma$'s decrease on the first
iteration of the map. However, this rule of thumb is much too
conservative --- there are large regions of this plot for which
one or more of the $\gamma$'s initially increase, and yet we are
below threshold.

It appears that distinguishing logical one-qubit gate errors from
logical two-qubit gate errors has an important quantitative effect
on our threshold estimates; the $\gamma_2$ curve turns upward much
more rapidly than the $\gamma_1$ curve if we are near but above
the threshold, and, in the vicinity of the fixed point in Figs.\
\ref{XXX} and \ref{quarter},
$\gamma_2$ is twice as large as
$\gamma_1$.  We see that this factor of two arises from a very
simple cause: the rectangle describing the replacement rule for
the two-qubit gate, Fig.\ \ref{2replace}, has two error correction
blocks that can fail.  Is this factor of two simply an artifact of
how we group the encoded computation into rectangles?  It is clear
that the answer is no; for the two-qubit gate, the key fact is
that the failure of either error-corrected block will cause the
entire encoded two-qubit gate, and the two encoded qubits emerging
from it, to be faulty.  It appears that this is the key reason
that the differing behavior of one- and two-qubit gates under
concatenation should be taken into account.

For memory errors, the story is rather different: we see that for
large parts of Fig.\ \ref{YYY} which are below threshold,
$\gamma_w$ increases (substantially, in fact) under concatenation.
This clearly arises from the fact that upon encoding, a waiting
period is replaced with an error correction step, with all its
(noisy) one- and two-qubit gates.  One might think, then, that it
might be desirable to skip error correction upon concatenation of
a memory location.  While this may indeed be possible, it raises a
danger that would require more extensive analysis to assess: since
single errors would go uncorrected, the error rate of qubits fed
into the following rectangle would be greater.  A much more
careful calculation of the effects of these passed-on errors would
need to be done to determine if skipping error corrections would
in fact be helpful.

Finally, we wish to note that the quantities $\alpha$ and $\beta$
are actually quite close to one near the threshold values of the
failure rates.  For $\beta$, we can understand this in the following
way: the probability of getting a nonzero syndrome, $1-\beta$, is
roughly the probability for a single fault among $N_{\rm syn}$
locations which make the syndrome nonzero, i.e.\ we can approximate
it as $N_{\rm syn}\gamma$. For this argument we forget about any
distinctions between types of errors and types of rectangles, so
the $N_{\rm syn}$ is some mean number of locations, and $\gamma$
is some average failure rate. Now, a rough estimate of the threshold
$\gamma$ (see Sec.\ \ref{threshestim}) is $1/{N \choose 2}$ where
$N$ is the number of locations that can cause errors on the data,
see Table \ref{failsourcetab}. When $N \sim N_{syn}$ which is the
case, we have $\beta\approx 1-2/N$. Since $N$ is somewhere between
100 and 200, we conclude that $\beta$ should be well above 90\%,
and this is what we see. A similar discussion can be given for
$\alpha$. In some cases, at the pseudothreshold, the values of
$\alpha$ and $\beta$ are much smaller.

\section{The Local Model With the $[[7,1,3]]$ Code}
\label{local7}

There are two main modifications that take place if we demand that
all gates be local.
 First, each error correction procedure needs to be modified so that it only consists
 of local gates. In this paper, we do not consider the additional overhead that is incurred from making the error correction
 local. Second, we have to use the local replacement rules as given in Figs.\ \ref{2replace} and \ref{wait1}--\ref{wait}.

The typical values for the scale factor $r$, which we will vary in
our numerical analysis, can be estimated by considering how many
qubits are in the error correction routine. For a nonlocal routine
this number of qubits (which includes one block of data qubits) is
 $k=7+2\times n_{rep}(7+3)$. In the regime (which we have found to be the relevant regime in the nonlocal numerical study)
 where $\alpha \rightarrow 1$, $n_{rep}\rightarrow 4$, this gives $k=87$. Note that we count both the ancillas
 in $\mathcal{X}$ and $\mathcal{Z}$ since the $\mathcal{X}$ ancillas will be prepared during the $\mathcal{Z}$ routine.
 By making the error correction local (for example by using dummy qubits) this number will increase somewhat. Thus it seems that
 taking $r$ in the range of 10--100 may be reasonable (for a two-dimensional architecture we may take $r \approx \lceil \sqrt{k} \rceil$
 which would give $r=10$). The operations that move qubits around over distance $r$ are composed from operations that move over distance $d$, where
 $r=\tau d$ and $\tau$ is some integer. We assume that the failure probability scales linearly with distance (which is a good assumption for
 small errors), i.e.\ if a ${\rm move(d)}$ operation has failure probability $\gamma_{md}$ then a ${\rm move(r)}$ operation has failure probability
 $\gamma_{mr}=\tau \gamma_{md}$.

As it turns out, in Steane's error-correcting procedure, there are
almost no one-qubit gates that occur in parallel with a two-qubit
gate. The only exception is the preparation of the verification
bits in the state $\ket{+}$ that occurs during $\mathcal{G}$, but
these can be prepared at the last convenient moment. This implies
that the computation is always a sequence of move gates followed
by local `in situ' gates. The modelling in Section \ref{reprules}
shows  that there are two types of wait locations, ones that
originally occur while a two-qubit gate occurs and ones that occur
during a one-qubit local gate. The wait locations of the first
type get mapped onto much longer wait and error correction
procedures, since they have to wait until the data has been moved.
We also assume that data has to be moved back in place for the
next gate, but it may be more efficient to move it elsewhere so
that it is ready for a possible next nonlocal gate.

In the upcoming analysis, we distinguish between the failure
probabilities for composite and elementary rectangles denoted as
$\vec{\gamma}^c(n)$ and $\vec{\gamma}^e(n)$. For $n=0$, we of
course have $\vec{\gamma}^c(0)=\vec{\gamma}^e(0)$. We enumerate
the types of locations and their probabilities in Table
\ref{local_loc}.

\begin{table}[h]
\begin{tabular}{|c|c|c|}
Location  &  Description & Failure Prob.\\ \hline\hline $1$ &
one-qubit gate  & $\gamma_1$ \\ \hline $2$ & two-qubit gate &
$\gamma_2$ \\ \hline $w1$ & wait during one-qubit gate &
$\gamma_{w1}$ \\ \hline $w2$ & wait during two-qubit gate &
$\gamma_{w2}$ \\ \hline $md$ & move distance d & $\gamma_{md}$
\\ \hline $wd$ & wait during ${\rm move(d)}$ & $\gamma_{wd}$ \\
\hline $1m$ & one-qubit gate $+$ measurement & $\gamma_{1m}$ \\
\hline $p$ & preparation & $\gamma_p$ \\ \hline
\end{tabular}
\caption{Types of locations and their failure probability symbols
in the local analysis.} \label{local_loc}
\end{table}

We now discuss the required modifications of the nonlocal model as
compared to the local analysis with the $[[7,1,3]]$ code.

\subsection{Modifications In The Failure Probability Estimation}

Each location $l$ gets replaced by a composite
 1-rectangle denoted as $R_l^c$ containing more than 1 elementary 1-rectangle, denoted as $R_j^e$. In order for the composite rectangle
 to fail at least one of the elementary rectangles has to fail, or
\beq \gamma_l^c(n)=1-\Pi_{j|j \in R^c_l} (1-\gamma_j^e(n)), \eeq
where the failure probabilities $\gamma_j^e(n)$ are calculated
similarly as in the nonlocal model (see Eqs. (\ref{1fail}) --
(\ref{2fail})).
 Table \ref{compelem} lists the occurrences of elementary 1-rectangles in composite 1-rectangles.
 The elementary failure probabilities $\gamma_j^e(n)$ are again functions of the vector of composite failure probabilities
 $\vec{\gamma}^c(n-1)$, i.e.\ $\gamma_j^e(n)={\bf F'}_j(\vec{\gamma}^c(n-1))$.

\begin{table}[h]
\begin{tabular}{|c|c|}
$l$ & $j\,|\,R_j^e \in R_l^c$ \\ \hline\hline
$1$ & $1[1]$ \\ \hline
$2$ & ${\rm move(d)}[2 \tau=2r/d]$, ${\rm wait(d)}[2 \tau]$, $2[1]$  \\ \hline
$1m$ & $1m[1]$ \\ \hline
$p$ & $p[1]$ \\ \hline
${\rm move(d)}$ & ${\rm move(d)}[r]$ \\ \hline
${\rm wait(d)}$ & ${\rm wait(d)}[r]$ \\ \hline
$w1$ &  $w1[1]$ \\ \hline
$w2$ &  ${\rm wait(d)}[2 \tau]$, $w2[1]$ \\\hline
\end{tabular}
\caption{Each location $l$ becomes a set of 1-rectangles by concatenation. The table lists which types of elementary 1-rectangles are present in the composite 1-rectangle $R_l^c$ based on the replacement rules of Figs.\ \protect\ref{2replace} and \protect\ref{wait1}--\protect\ref{wait}. The number between $[]$ indicates how often the elementary 1-rectangle occurs inside the composite 1-rectangle.}
\label{compelem}
\end{table}

Now we list the necessary modifications to the failure probability of an elementary rectangle and the estimation of $\alpha$, the probability of an ancilla passing verification, and $\beta$,  the probability of obtaining a zero syndrome. Note
 that the failure probability is now a function of the composite failure probabilities at the lower level.
 First we list the modifications to Table \ref{failsourcetab} in Table \ref{modfailsourcetab}. In the source
 `errors due to propagation from the ancilla', we also need to use a modified $\alpha$, $\beta$, and $\mathbb{P}(\mbox{pass and no $X$})$, estimated in the
 next section.

\begin{table}[h]
\begin{tabular}{|c|c|c|}
                       Modified Source                                      & $\delta$  &   $N$   \\ \hline\hline
{\small Memory faults on data at the end of $\mathcal{S}$}
& $\gamma_{w1}^c$ & $14(s_x+s_z)$ \\ \hline {\small Memory faults
on data during $\mathcal{R}$}                  & $\gamma_{w1}^c$ &
$6(\delta_{s_z,s}+\delta_{s_x,s})$ \\ \hline {\small Memory faults
($w1$) on data when $s=1$}                            &
$\gamma_{w1}^c$ & $14(s-1)(\delta_{s_z,1}+\delta_{s_x,1})$\\
\hline {\small Memory faults ($w2$) on data when $s=1$}
& $\gamma_{w2}^c$ & $7(s-1)(\delta_{s_z,1}+\delta_{s_x,1})$\\
\hline {\small $X$ errors on ancillas waiting ($w1$) for
$\mathcal{S}$}              & $\gamma_{w1}^c$ & $14
s(s-1)(\delta_{s_z,s}+\delta_{s_x,s})/2$ \\ \hline {\small $X$
errors on ancillas waiting ($w2$) for $\mathcal{S}$}
& $\gamma_{w2}^c$ & $7 s(s-1)(\delta_{s_z,s}+\delta_{s_x,s})/2$ \\
\hline
\end{tabular}
\caption{Modified memory sources of failure and their contribution to the failure probability. We only list the sources that are different due to the distinction between $w1$ and $w2$, the other sources are unchanged.}
\label{modfailsourcetab}
\end{table}

For a rectangle that acts on a single block, i.e.\ $l=p,w1,w2,1,1m,{\rm move(d)},{\rm wait(d)}$, we write, similar to Eq.\ (\ref{1fail})
\beq
\gamma_l^e(n)=\beta^2 {\bf F'}_l[1,1](\vec{\gamma}^c(n-1))
+2\beta(1-\beta){\bf F'}_l[s,1](\vec{\gamma}^c(n-1))+(1-\beta)^2 {\bf F'}_l[s,s](\vec{\gamma}^c(n-1)),
\label{mod1fail}
\eeq
where the function ${\bf F'}_l$ takes into account the modifications in the failure sources.


\subsection{Modifications in $\alpha$ and $\beta$}

In each of the expressions in Section \ref{estmab} (see Eqns.\ (\ref{zanc}) -- (\ref{waitdat})) , we have to use the failure probabilities of the {\em composite} rectangles.
Equations (\ref{zanc}) and (\ref{xanc}) change due to the distinction between $w1$ and $w2$ locations:
\bea
\mathbb{P}(\mbox{pass and no $Z$})=(1-\gamma_p^c)^4 (1-\gamma_1^c)^4 (1-2\gamma_{1m}^c/3)^4 \times \nonumber \\
\Pi_{i \in \mathcal{G}}(1-\gamma_i^c)^{N(i \in
\mathcal{G})}(1-2\gamma_{w2}^c/3)^{12} (1-2\gamma_{w1}^c/3)^{14}
(1-\gamma_{w2}^c)^6 (1-\gamma_2^c)^{13}, \label{newzanc} \eea
and, slightly different,
\bea
\mathbb{P}(\mbox{pass and no $X$})=(1-2\gamma_p^c/3)^4 (1-2\gamma_1^c/3)^4 (1-2\gamma_{1m}^c/3)^4 \times \nonumber \\
\Pi_{i \in \mathcal{G}}(1-2\gamma_i^c/3)^{N(i \in
\mathcal{G})}(1-2\gamma_{w2}^c/3)^{18} (1-2\gamma_{w1}^c/3)^{14}
(1-\gamma_2^c)^{13}, \label{newxanc} \eea

Note that in Eq.\ (\ref{ince}) we maximize over all possible locations in this new model.
We also distinguish between $w1$ and $w2$ in Eq.\ (\ref{waitdat}):
\beq
\mathbb{P}(\mbox{no $X$ err. on waiting data})=(1-2\gamma_{w1}^c/3)^{14(s-1)}(1-2\gamma_{w2}^c/3)^{7(s-1)}.
\label{newwaitdat}
\eeq

\section{Numerical Threshold Studies for The Local Model}

By numerical iteration of the equations of the preceding sections,
we study the repeated application of the map determined by
encoding with the $[[7,1,3]]$ code in the local model.  Although we now have
an even higher-dimensional map than in the nonlocal studies
(eight dimensions rather than five), the two cases are
mathematically very similar; it is evident that the structure of
the flows is again determined by the presence of an unstable fixed
point with one positive eigenvalue (and in this case $8-1=7$
negative eigenvalues).  An important difference is that the local
map contains a free parameter, $\tau$, the frequency of error
correction while moving; we will exploit the freedom to optimize
the fault-tolerance threshold in the numerical studies below.

\begin{figure}[htb]
\begin{center}
\epsfxsize=15cm \epsffile{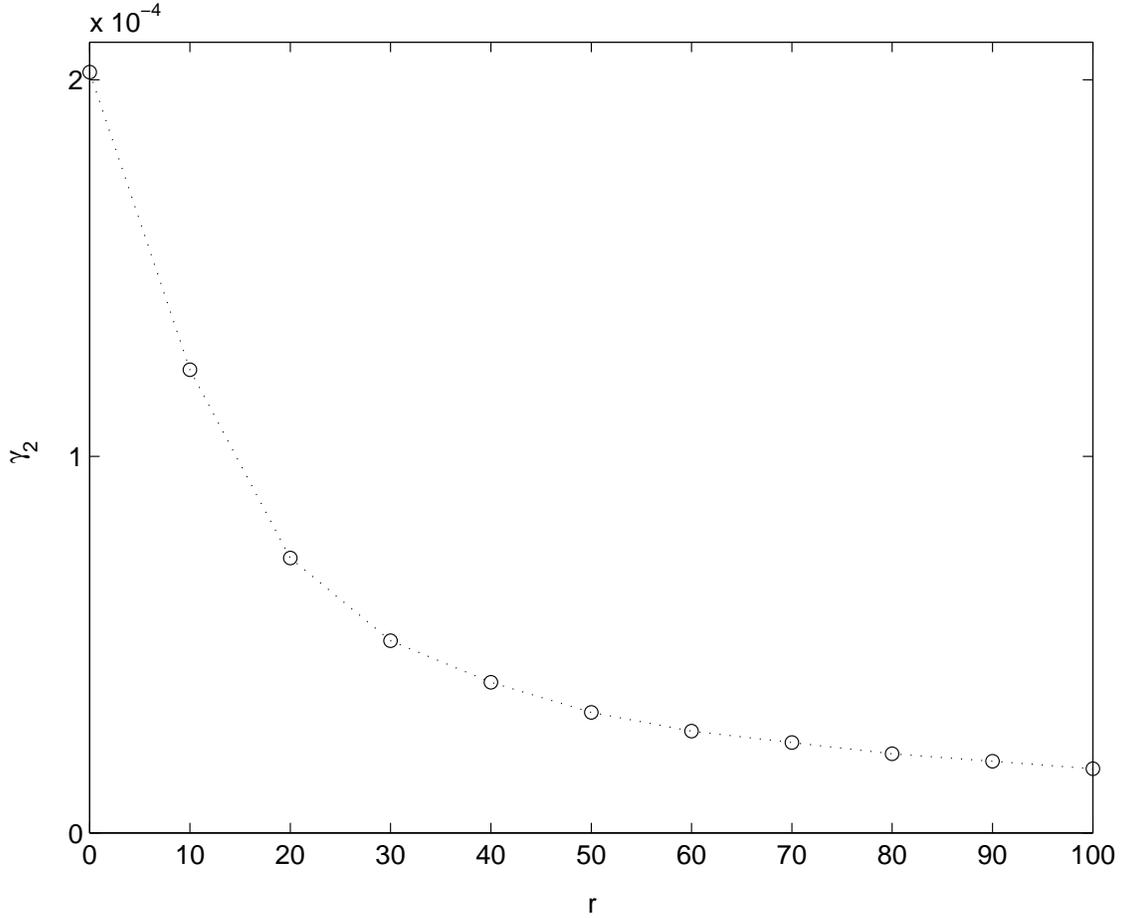} \caption{Gate
failure rate threshold versus the scale parameter $r$ for the
local model. We have taken $\gamma_1=\gamma_2=\gamma_m=\gamma_p$,
$\gamma_{w1}=\gamma_{w2}=0.1\times\gamma_2$,
$\gamma_{wd}=0.1\times\gamma_{md}$, and
$\gamma_{md}=r/\tau\times\gamma_2$.  $\tau$, the frequency with
which a qubit is error-corrected while being moved over distance r
is optimized in every case.  The threshold follows very close to a
$1/r$ dependence.} \label{ZZZ}
\end{center}
\end{figure}
Our first observation, illustrated in Fig.\ \ref{ZZZ}, is that the
numerical values of the threshold failure probabilities can in fact
be strongly affected by the need to transport qubits.  For this
figure we take physical failure rates
$\gamma_1=\gamma_2=\gamma_m=\gamma_p$,
$\gamma_{w1}=\gamma_{w2}=0.1\times\gamma_2$,
$\gamma_{wd}=0.1\times\gamma_{md}$, and
$\gamma_{md}=r/\tau\times\gamma_2=d \times \gamma_2$.  In words,
this means that the gate, measurement, and preparation failure rates
are taken all equal, wait errors (per unit time or per unit
distance travelled during moving periods) are one-tenth of the gate
failure rate, and moving a qubit over a unit distance is as noisy as
a gate operation (corresponding to a scenario, say, in which
moving over unit distance requires an actual swap gate execution).
We have also optimized $\tau$ to be $\tau=4$, that is, $d = \lceil(r/\tau)\rceil = 13$, which
means that error correction is performed on qubits in transit once
every 13 units of distance moved (13 swap gates, say).

\begin{figure}[htb]
\begin{center}
\epsfxsize=15cm \epsffile{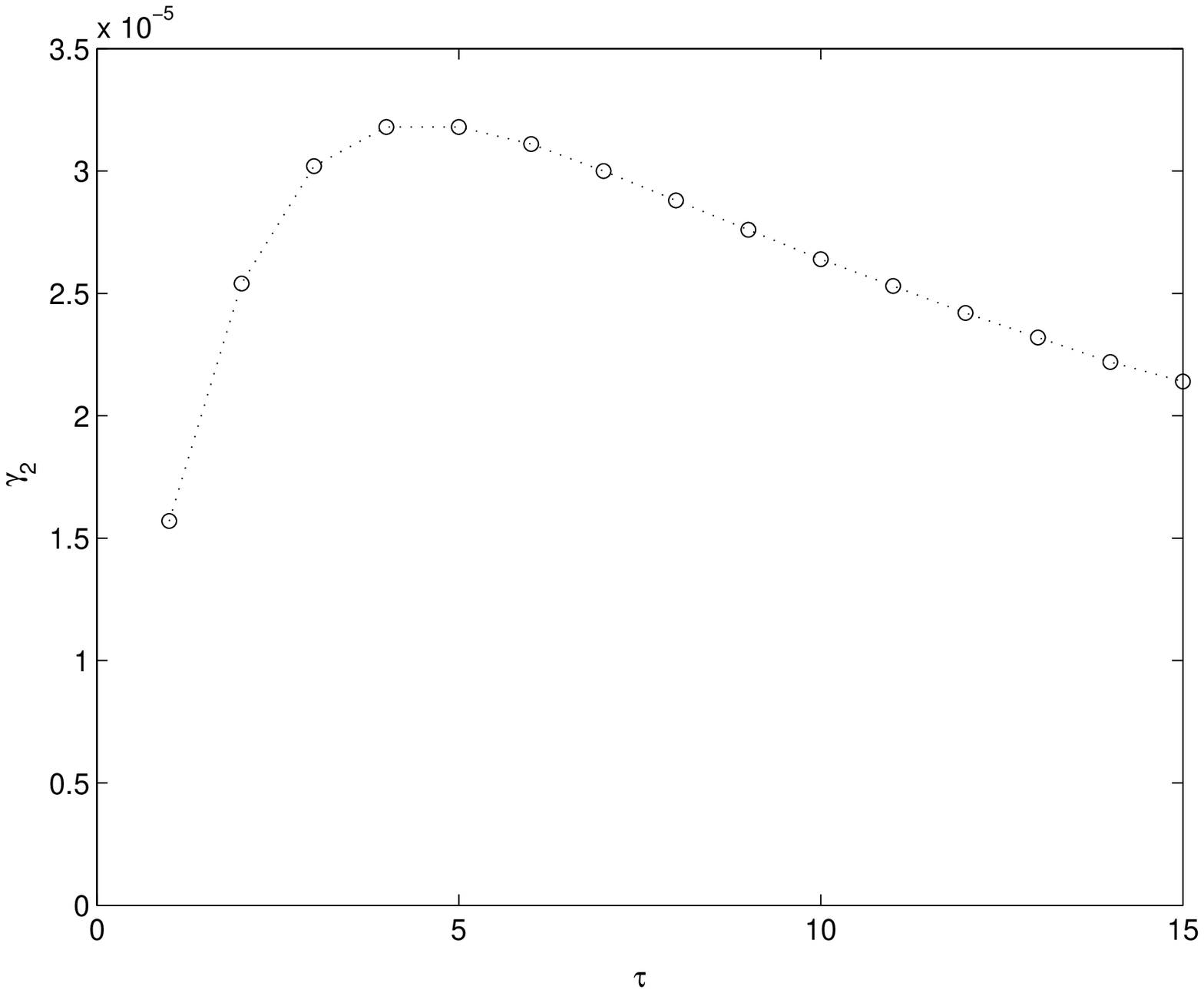}
\caption{Gate failure rate threshold versus $\tau$, the frequency of error correction of a transported qubit, for $r=50$.
As in Fig.\ \protect{\ref{ZZZ}}, we have taken $\gamma_1=\gamma_2=\gamma_m=\gamma_p$, $\gamma_{w1}=\gamma_{w2}=0.1\times\gamma_2$, $\gamma_{wd}=0.1\times\gamma_{md}$, and $\gamma_{md}=r/\tau\times\gamma_2$.  While not very strongly $\tau$-dependent, the optimal threshold occurs at $\tau=4$.}
\label{AAA}
\end{center}
\end{figure}

\begin{figure}[htb]
\begin{center}
\epsfxsize=15cm \epsffile{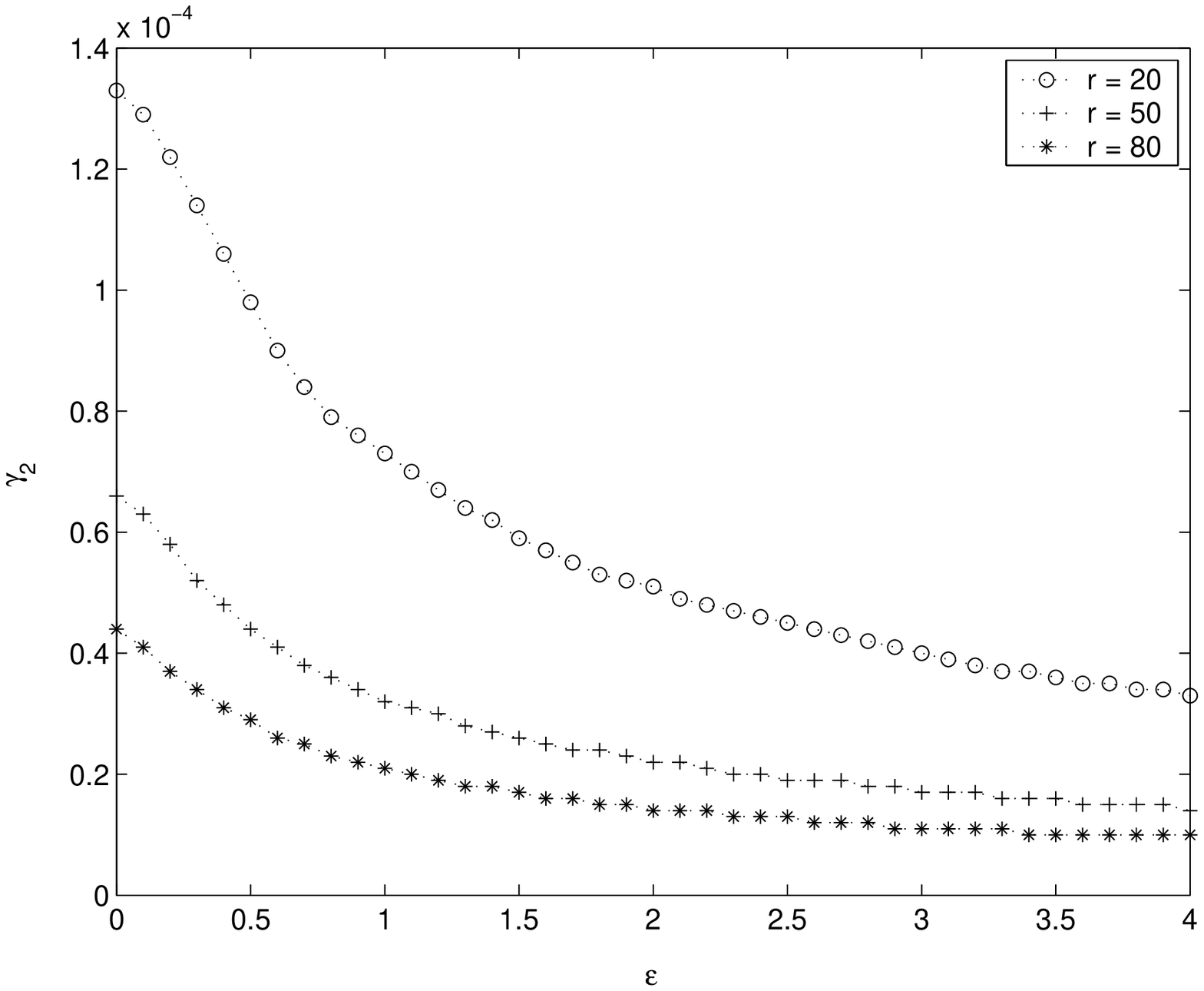}
\caption{Gate failure rate threshold versus $\epsilon$, a parameter that measures the relative noise rate per unit distance for a qubit being moved.
We initially set $\gamma_1=\gamma_2=\gamma_m=\gamma_p$, $\gamma_{w1}=\gamma_{w2}=0.1\times\gamma_2$, $\gamma_{wd}=0.1\times r/\tau\times\gamma_2$, and $\gamma_{md}=\epsilon  r/\tau\times\gamma_2$.
Scale parameters $r$ equals 20, 50, and 80 are studied.  At every point $\tau$ is re-optimized. The dependence on $\epsilon$ is slow, evidently slower than $1/\epsilon$.}
\label{BBB}
\end{center}
\end{figure}

As Fig.\ \ref{ZZZ} shows, for these conditions the threshold (we
plot the $\gamma_2$ threshold value) decreases strongly with $r$;
the dependence is very close to $\gamma_2^{\rm thresh}\propto
1/r$, confirming the analysis in Section \ref{threshestim}. Note however
findings are more optimistic here than the analytical lower bound in
Section \ref{threshestim}, that is, we see that
$\gamma_2^{\rm thresh,loc} \approx \gamma_2^{\rm thresh, nonloc} \times c/r$ for
some constant $c$ which is a bit larger than 1. For a
scale parameter $r=20$, which could well be a reasonable number,
we get $\gamma_2^{\rm thresh.}=0.73\times 10^{-4}$, nearly an
order of magnitude below the numbers typical in the nonlocal
model, shown in Fig.\ \ref{YYY}. We have plotted these results in the high
noise limit, but we have found similar behavior when the noise
during transit is not very high, as seen in the dependence on $r$ in Fig.\ \ref{BBB} for small $\epsilon$.

Fig.\ \ref{AAA} shows the result of varying $\tau$ for the failure
probability choices of Fig.\ \ref{ZZZ}, with fixed $r=50$. We do
this by choosing a $\tau$ that minimizes the threshold probability.
After that we fix $\tau$ to be the optimal value, that is we do
not adjust $\tau$ at each level of concatenation. While the
threshold value is not a very strong function of $\tau$, it is
clearly optimal for $\tau=4$. In more general studies in which we
vary the initial values for $\gamma$ and $r$ we do not find a
simple relation between the optimal $\tau$ and these parameters.

This result was initially surprising to us, since it says that it
is optimal to allow the moving qubits to become about thirteen
times noisier than the qubits involved in gate operations before
they are error corrected.  The explanation for this seems to be
that since qubits in motion do not have a chance of spreading
error to other data qubits, allowing them to get noisier is not
dangerous, and is actually desirable given the level of errors
introduced by the error correction step itself. Of course, before
they couple to other qubits we perform an error-correcting step in
order to get rid of the accumulation of errors. A similar choice
of less frequent error correction may be advantageous for a qubit
who undergoes a few one-qubit and wait locations in succession. In
such a case, errors do not spread to other blocks during these
procedures and we finish the sequence by an error-correcting step
as in the qubits in transit case.

Finally, Fig.\ \ref{BBB} shows the result of varying between
noiseless moving and high-noise moving scenarios.  This is
captured by varying the parameter $\epsilon$ in the setting
$\gamma_{md}=\epsilon r/\tau\times\gamma_2$.  The choice for
$\gamma_{md}$ reflects the idea that the failure rates for qubits
that are waiting during a move step should depend only on the
distance moved (and therefore, the time waiting during each
elementary move step).  $\epsilon=1$ is exactly the scenario
explored in Figs. \ref{ZZZ} and \ref{AAA}. $\epsilon=0$
corresponds to free moving, in which the qubit can be converted
into some noiseless flying form for transportation; a rather
artificial feature of this limit is that waiting is then noisier
than moving.
In Fig.\ \ref{BBB}, the other parameters are initially set as:
$\gamma_1=\gamma_2=\gamma_m=\gamma_p$,
$\gamma_{w1}=\gamma_{w2}=0.1\times\gamma_2$, and
$\gamma_{wd}=0.1\times r /\tau\times\gamma_2$.


It is evident from this that the error threshold is a weaker
function of the moving failure rate $\epsilon$ than it is of the
scale parameter $r$. When $\epsilon \rightarrow 0$ the waiting
during moving is more error-prone than the moving itself and this
waiting should be the main cause of the $1/r$ behavior in this
limit. In other words it is the scale-up of the circuit with every
level of concatenation and the additional waiting this causes,
i.e.\ a ${\rm wait(d)}$ location gets replaced by $r$ ${\rm
wait(d)}$ locations, that is the dominant reason why the threshold
is lower than in the nonlocal model.

On the other hand, the ``weak'' dependence on $\epsilon$ seems to
indicate that repeated error correction during moving is able to
maintain acceptable fidelity for the moved qubits even in the face
of moving errors.


This gives some new hope for schemes, such as those involving
spins in semiconductors or Josephson junctions, in which qubit
moving is inherently as difficult as gate operations. We know that
in such a high failure-rate regime, entanglement distribution
followed by purification and then teleportation, can be a more
effective way of moving qubits \cite{knill:ftps, reichardt:ft,bennett+:bdsw}. The rather
strong sensitivity to $r$ that we find (Fig.\ \ref{ZZZ}) suggests
that if such strategies are employed, they should best be used in
a way which does not increase the number of ancillas needed, and
hence the scale parameter, too much.

Our numerics of course add a note of caution to this optimism:
although the $\epsilon$ dependence we find is not too severe, over
most of the range of the plot in Fig.\ \ref{BBB}, the actual values
of the fault-tolerance threshold failure rate is well below
$10^{-4}$, in a range that is presently far, far beyond the
capability of any quantum computer prototype in the laboratory.

\section{Outlook}

We see at least two extensions of this direction of research. One is to indeed make the error correction routine local, assuming some
 mechanism for short-distance transportation and a spatial layout of the qubits. We could then redo our local analysis, possibly with
 some more lengthy analysis of the failure probability that includes more details, in order to get a full estimate of the change in
 threshold due to locality. Secondly, one needs to consider where all the additional error correction in transit and moving will take
 place and has to design a layout for this. Given this layout there may be modifications to the replacement rules in order to reflect the real
 architecture.

\section{Acknowledgements}
Our quantum circuit diagrams were made using the Q-circuit \LaTeX\ macro package by Steve Flammia and Bryan Eastin.
Krysta Svore acknowledges support by the NPSC.
Barbara Terhal and David DiVincenzo acknowledge
support by the NSA and the ARDA through ARO contract numbers DAAD19-01-C-0056
and W911NF-04-C-0098.

\appendix

\section{Replacement Rules}

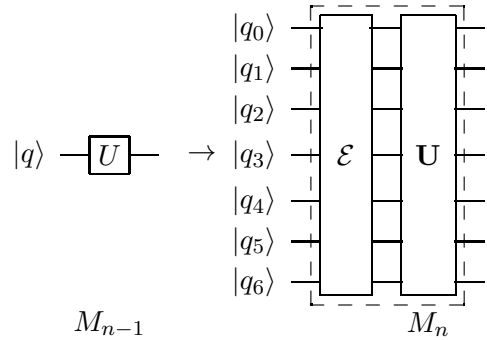
\begin{figure}[htbp]
\mbox{
\Qcircuit @C=1em @R=.5em {
      & & & & & \lstick{\ket{q_0}} & \multigate{6}{\mathcal E} &
\multigate{6}{\bf U} &
\qw \\
    & & & & & \lstick{\ket{q_1}} & \ghost{E} &
\ghost{G} &
\qw \\
    & & & & & \lstick{\ket{q_2}} & \ghost{E} &
\ghost{G} &
\qw \\
    \lstick{\ket{q}} & \gate{U} & \qw & \push{\rightarrow\rule{.3em}{0em}}& & \lstick{\ket{q_3}} & \ghost{E} &
\ghost{G} &
\qw \\
    & & & & & \lstick{\ket{q_4}} & \ghost{E} &
\ghost{G}&
\qw \\
    & & & & & \lstick{\ket{q_5}} & \ghost{E} &
\ghost{G} &
\qw \\
    & & & & & \lstick{\ket{q_6}} & \ghost{E} &
\ghost{G}&
\qw \gategroup{1}{7}{7}{8}{.6em}{--}
\\
    & & & & & & & & & & & &\\
    & \mbox{$M_{n-1}$} &
& & & & & \mbox{$M_n$}& & & & }} \caption{The replacement rule for
a one-qubit gate location $U$ or a ${\rm wait1}$ location. The
dashed box represents a 1-rectangle.  $\mathcal E$ represents the
error correction procedure. ${\bf U}$ represents the local
fault-tolerant implementation of $U$. Note that in each figure, a
qubit in $M_{n-1}$ is encoded as $m=7$ qubits in $M_n$.}
\label{wait1}
\end{figure}

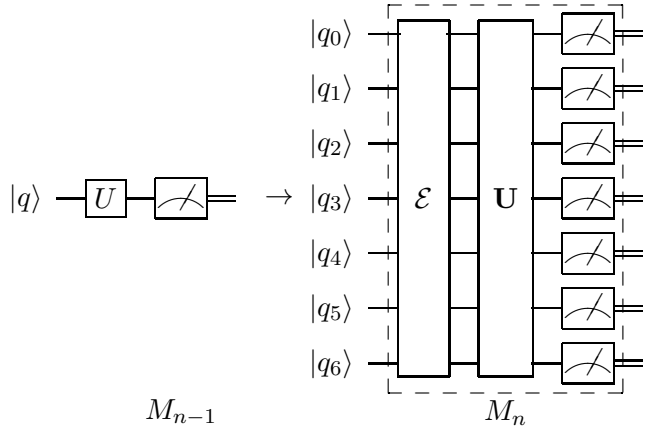
\begin{figure}[htbp]
\mbox{
\Qcircuit @C=1em @R=.5em {
      & & & & & & \lstick{\ket{q_0}} &
    \multigate{6}{\mathcal E} &
\multigate{6}{\bf U} &
\meter & \cw \\
    & & & & & & \lstick{\ket{q_1}} & \ghost{E} &
\ghost{G} &
 \meter & \cw \\
    & & & & & & \lstick{\ket{q_2}} & \ghost{E} &
\ghost{G}&
\meter & \cw \\
    \lstick{\ket{q}} & \gate{U} & \meter & \cw & \push{\rightarrow\rule{.3em}{0em}}& & \lstick{\ket{q_3}} & \ghost{E} &
\ghost{G}&
\meter & \cw \\
    & & & & & & \lstick{\ket{q_4}} & \ghost{E} &
\ghost{G} &
\meter & \cw \\
    & & & & & & \lstick{\ket{q_5}} & \ghost{E} &
\ghost{G}&
\meter & \cw \\
    & & & & & & \lstick{\ket{q_6}} & \ghost{E} &
\ghost{G}&
\meter & \cw \gategroup{1}{8}{7}{10}{.6em}{--}
\\
    & & & & & & & & & &\\
    & & \mbox{$M_{n-1}$} & & & & & & \mbox{$M_n$}& & & & &
} }
\caption{The replacement rule for a one-qubit gate $U$ followed by a measurement.}
\label{1mreplace}
\end{figure}

\begin{figure}[htbp]
\mbox{
\Qcircuit @C=1em @R=.5em {
    & & & & & \lstick{\ket{q_0}} & \multigate{6}{\mathcal E} & \gate{{\rm wait(r)}} & \multigate{6}{\mathcal E} & \gate{\rm wait2} & \multigate{6}{\mathcal E} & \gate{{\rm wait(r)}} & \qw \\
    & & & & & \lstick{\ket{q_1}} & \ghost{E} & \gate{{\rm wait(r)}} & \ghost{E} & \gate{\rm wait2} & \ghost{E} & \gate{{\rm wait(r)}} & \qw \\
    & & & & & \lstick{\ket{q_2}} & \ghost{E} & \gate{{\rm wait(r)}} & \ghost{E} & \gate{\rm wait2} & \ghost{E} & \gate{{\rm wait(r)}} & \qw \\
    \lstick{\ket{q}} & \gate{\rm wait2} & \qw & \push{\rightarrow\rule{.3em}{0em}} & & \lstick{\ket{q_3}} & \ghost{E} & \gate{{\rm wait(r)}} & \ghost{E} & \gate{\rm wait2} & \ghost{E} & \gate{{\rm wait(r)}} & \qw \\
    & & & & & \lstick{\ket{q_4}} & \ghost{E} & \gate{{\rm wait(r)}} & \ghost{E} & \gate{\rm wait2} & \ghost{E} & \gate{{\rm wait(r)}} & \qw \\
    & & & & & \lstick{\ket{q_5}} & \ghost{E} & \gate{{\rm wait(r)}} & \ghost{E} & \gate{\rm wait2} & \ghost{E} & \gate{{\rm wait(r)}} & \qw \\
    & & & & & \lstick{\ket{q_6}} & \ghost{E} & \gate{{\rm wait(r})} & \ghost{E} & \gate{\rm wait2} & \ghost{E} & \gate{{\rm wait(r)}} & \qw \gategroup{1}{7}{7}{8}{.6em}{--} \gategroup{1}{9}{7}{10}{.6em}{--} \gategroup{1}{11}{7}{12}{.6em}{--}\\
    & & & & & & & & & & & & \\
    & \mbox{$M_{n-1}$} & & & & & & & & \mbox{$M_n$} & & &
} } \caption{The replacement rule for a ${\rm wait2}$ (also called
$w2$) location acting in parallel with a two-qubit gate. The
replacement circuit contains three elementary 1-rectangles.}
\label{wait2}
\end{figure}
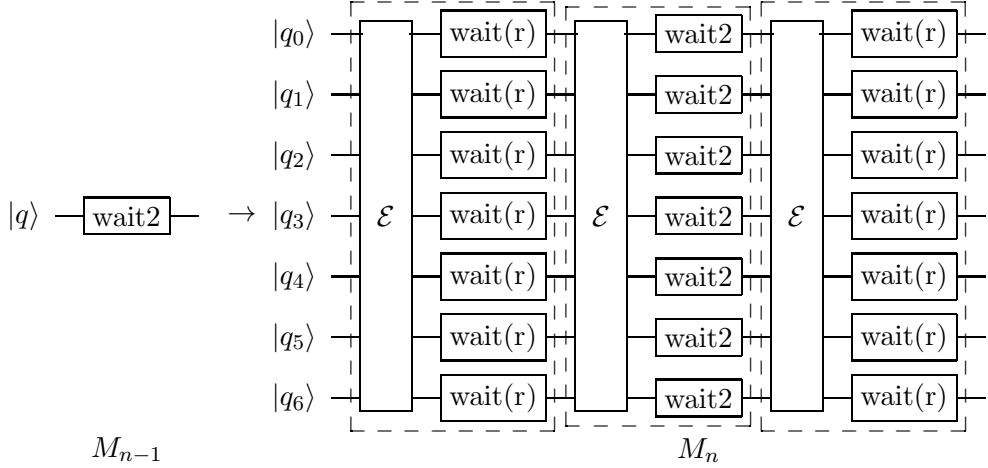

\begin{figure}[tbp]
\mbox{
\Qcircuit @C=1em @R=.5em {
    & & & & & \lstick{\ket{q_0}} & \multigate{6}{\mathcal E} & \gate{{\rm move(r)}}&\qw & & & \multigate{6}{\mathcal E} & \gate{{\rm move(r)}} & \qw \\
    & & & & & \lstick{\ket{q_1}} & \ghost{E} & \gate{{\rm move(r)}}&\qw & & & \ghost{E} & \gate{{\rm move(r)}} & \qw \\
    & & & & & \lstick{\ket{q_2}} & \ghost{E} & \gate{{\rm move(r)}}&\qw & & & \ghost{E} & \gate{{\rm move(r)}} & \qw \\
    \lstick{\ket{q}} & \gate{{\rm move(r)}} & \qw & \push{\rightarrow\rule{.3em}{0em}} & & \lstick{\ket{q_3}} & \ghost{E} & \gate{{\rm move(r)}}&\qw & \push{...\rule{.3em}{0em}}& & \ghost{E} & \gate{{\rm move(r)}} & \qw \\
    & & & & & \lstick{\ket{q_4}} & \ghost{E} & \gate{{\rm move(r)}}&\qw & & & \ghost{E} & \gate{{\rm move(r)}} & \qw \\
    & & & & & \lstick{\ket{q_5}} & \ghost{E} & \gate{{\rm move(r)}}&\qw & & & \ghost{E} & \gate{{\rm move(r)}} & \qw \\
    & & & & & \lstick{\ket{q_6}} & \ghost{E} & \gate{{\rm move(r)}}&\qw & & & \ghost{E} & \gate{{\rm move(r)}} & \qw \gategroup{1}{7}{7}{8}{.6em}{--} \gategroup{1}{12}{7}{13}{.6em}{--} \\
    & & & & & & & & & & & & &\\
    & \mbox{$M_{n-1}$} & & & & & & & & \mbox{$M_n$}& & & &
} }
\caption{The replacement rule for a ${\rm move(r)}$ gate. The replacement circuit contains $r$ elementary 1-rectangles.}
\label{move}
\end{figure}
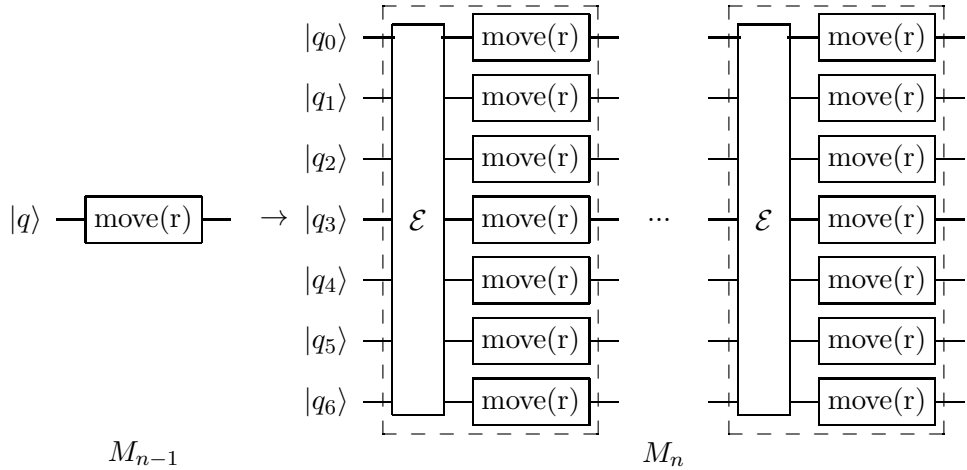

\begin{figure}[tbp]
\mbox{
\Qcircuit @C=1em @R=.5em {
    & & & & & \lstick{\ket{q_0}} & \multigate{6}{\mathcal E} & \gate{{\rm wait(r)}}&\qw & & & \multigate{6}{\mathcal E} & \gate{{\rm wait(r)}} & \qw \\
    & & & & & \lstick{\ket{q_1}} & \ghost{E} & \gate{{\rm wait(r)}}&\qw & & & \ghost{E} & \gate{{\rm wait(r)}} & \qw \\
    & & & & & \lstick{\ket{q_2}} & \ghost{E} & \gate{{\rm wait(r)}}&\qw & & & \ghost{E} & \gate{{\rm wait(r)}} & \qw \\
    \lstick{\ket{q}} & \gate{{\rm wait(r)}} & \qw & \push{\rightarrow\rule{.3em}{0em}} & & \lstick{\ket{q_3}} & \ghost{E} & \gate{{\rm wait(r)}}&\qw & \push{...\rule{.3em}{0em}}& & \ghost{E} & \gate{{\rm wait(r)}} & \qw \\
    & & & & & \lstick{\ket{q_4}} & \ghost{E} & \gate{{\rm wait(r)}}&\qw & & & \ghost{E} & \gate{{\rm wait(r)}} & \qw \\
    & & & & & \lstick{\ket{q_5}} & \ghost{E} & \gate{{\rm wait(r)}}&\qw & & & \ghost{E} & \gate{{\rm wait(r)}} & \qw \\
    & & & & & \lstick{\ket{q_6}} & \ghost{E} & \gate{{\rm wait(r)}}&\qw & & & \ghost{E} & \gate{{\rm wait(r)}} & \qw \gategroup{1}{7}{7}{8}{.6em}{--} \gategroup{1}{12}{7}{13}{.6em}{--} \\
    & & & & & & & & & & & & &\\
    & \mbox{$M_{n-1}$} & & & & & & & & \mbox{$M_n$}& & & &
} }
\caption{The replacement rule for a ${\rm wait(r)}$ gate. The replacement circuit contains $r$ elementary 1-rectangles.}
\label{wait}
\end{figure}
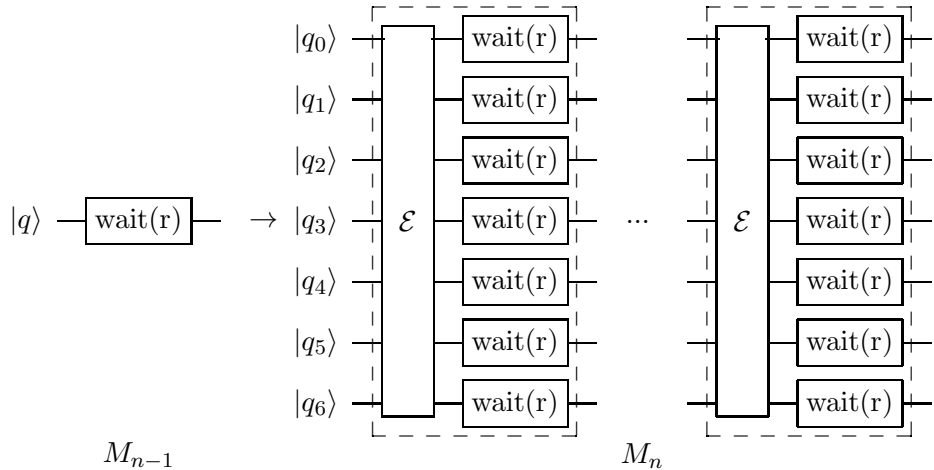

\newpage

\section{Definitions Of $n$-Rectangles, Blocks And Sparseness}\label{defft}

\begin{itemize}
\item
A set of qubits in $M_n$ is called an {\it $s$-block} if they
 originate from one qubit in $M_{n-s}$.  A {\it $s$-rectangle} in $M_n$
 is a set of locations that originates from one location in $M_{n-s}$.
 A {\it $s$-working period} is the time interval in $M_n$ which
 corresponds to one time step in $M_{n-s}$.
\item
Let $B$ be a set of $n$-blocks in the computation $M_n$.  An
 {\it $(n,k)$-sparse} set of qubits $A$ in $B$ is a set of qubits in which for every $n$-block in $B$,
 there are at most $k$ $(n-1)$-blocks such that the set $A$ in this block
 is not $(n-1,k)$-sparse.  A $(0,k)$-sparse set of qubits is an empty
 set of qubits.
\item A set of locations in a $n$-rectangle is {\it $(n,k)$-sparse} when there are
 at most $k$ $(n-1)$-rectangles such that the set is not $(n-1,k)$-sparse
  in that $(n-1)$-rectangle.  A $(0,k)$-sparse set of locations in a
 $0$-rectangle is an empty set.  A fault-path in $M_n$ is $(n,k)$-sparse
 if in each $n$-rectangle, the set of faulty locations is $(n,k)$-sparse.
 \item A computation code $C$ has {\it spread} $t$ if one fault occurring in a
 particular $1$-working period affects at most $t$ qubits in each
 1-block, i.e.\ causes at most $t$ errors in each 1-block in that particular
 working period.
\end{itemize}

\newpage

\section{Error-correcting using the $[[7,1,3]]$ Code}\label{ecor}

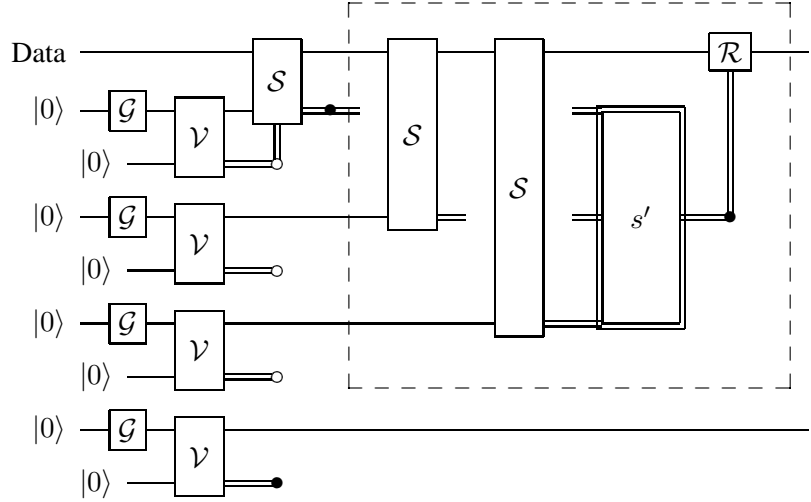
\begin{figure}[hb]
\mbox{ \Qcircuit @C=1em @R=.7em {
&&&&&&&&&&&&&&&\\
&\lstick{\mbox{Data}}&\qw&\qw&\multigate{1}{\mathcal{S}}&\qw&\qw&\multigate{3}{\mathcal{S}}&\qw&\multigate{5}{\mathcal{S}}&\qw&\qw&\qw&\qw&\gate{\mathcal{R}}&\qw&\qw\\
&\lstick{\ket{0}}
&\gate{\mathcal{G}}&\multigate{1}{\mathcal{V}}&\ghost{\mathcal{S}}&\control
\cw&\cw&&&&&\cw&&&&\\
&&\lstick{\ket{0}}&\ghost{\mathcal{V}}&\controlo \cw\cwx&&&&&&&&&&&\\
&\lstick{\ket{0}}
&\gate{\mathcal{G}}&\multigate{1}{\mathcal{V}}&\qw&\qw&\qw&\ghost{\mathcal{S}}&\cw&&&\cw&\push{s'}&&\control\cw\cwx[-3]&\\
&&\lstick{\ket{0}} &\ghost{\mathcal{V}}&\controlo\cw&&&&&&&&&&&\\
&\lstick{\ket{0}}
&\gate{\mathcal{G}}&\multigate{1}{\mathcal{V}}&\qw&\qw&\qw&\qw&\qw&\ghost{\mathcal{S}}&\cw&\cw&&&&\\
&&\lstick{\ket{0}} &\ghost{\mathcal{V}}&\controlo\cw&&&&&&&&&&&\\
&\lstick{\ket{0}}
&\gate{\mathcal{G}}&\multigate{1}{\mathcal{V}}&\qw&\qw&\qw&\qw&\qw&\qw&\qw&\qw&\qw&\qw&\qw&\qw&\qw\\
&&\lstick{\ket{0}}&\ghost{\mathcal{V}}&\control\cw
\gategroup{3}{12}{7}{14}{.3em}{=}
\gategroup{1}{7}{8}{16}{.7em}{--}
}}
\caption{The Steane $X$-error correction protocol, $\mathcal X$ \protect{\cite{steane:overhead}}.  The black circle represents control on a nonzero result.  A white circle represents control on a zero result. $s'$ represents a classical procedure to check if $s'$ of the $s$ syndromes agree. The dashed box procedure is applied only if the controlling syndrome is not zero.  There are $n_{rep}$ prepared ancilla blocks. Each line represents 7 qubits.  After $\mathcal{V}$, $\alpha$ `good' verification blocks remain. $\mathcal{R}$ represents the recovery procedure.}
\label{QEC}
\end{figure}

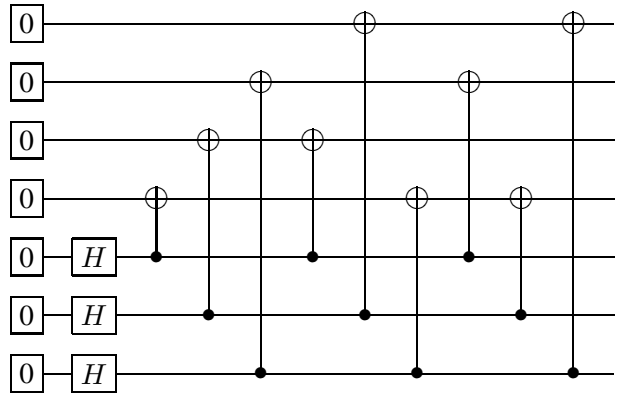
\begin{figure}[h]
\mbox{ \Qcircuit @C=1em @R=.7em {
\push{\fbox{0}\rule{0em}{0em}}&\qw&\qw&\qw&\qw&\qw&\targ&\qw&\qw&\qw&\targ
&\qw\\
\push{\fbox{0}\rule{0em}{0em}}&\qw&\qw&\qw&\targ&\qw&\qw&\qw&\targ&\qw&\qw
&\qw\\
\push{\fbox{0}\rule{0em}{0em}}&\qw&\qw&\targ&\qw&\targ&\qw&\qw
&\qw&\qw&\qw
&\qw\\
\push{\fbox{0}\rule{0em}{0em}}&\qw&\targ&\qw &\qw&\qw&\qw
&\targ&\qw&\targ&\qw&\qw\\
\push{\fbox{0}\rule{0em}{0em}}&\gate{H}&\ctrl{-1}&\qw&\qw
&\ctrl{-2}&\qw&\qw &\ctrl{-3}&\qw&\qw
&\qw\\
\push{\fbox{0}\rule{0em}{0em}}&\gate{H}&\qw&\ctrl{-3}&\qw
&\qw&\ctrl{-5}&\qw & \qw&\ctrl{-2}&\qw
&\qw\\
\push{\fbox{0}\rule{0em}{0em}}& \gate{H}&\qw&\qw&\ctrl{-5}
&\qw&\qw&\ctrl{-3} &\qw&\qw&\ctrl{-6}
&\qw
}} \caption{The $\mathcal{G}$ network for $X$ or $Z$-error
correction \protect{\cite{preskill:faulttol}}.  The network can be
executed in 5 time steps.  It produces the encoded $|0\rangle$
state.  The boxed zero represents preparation of a $|0\rangle$
state.} \label{G_parallel}
\end{figure}

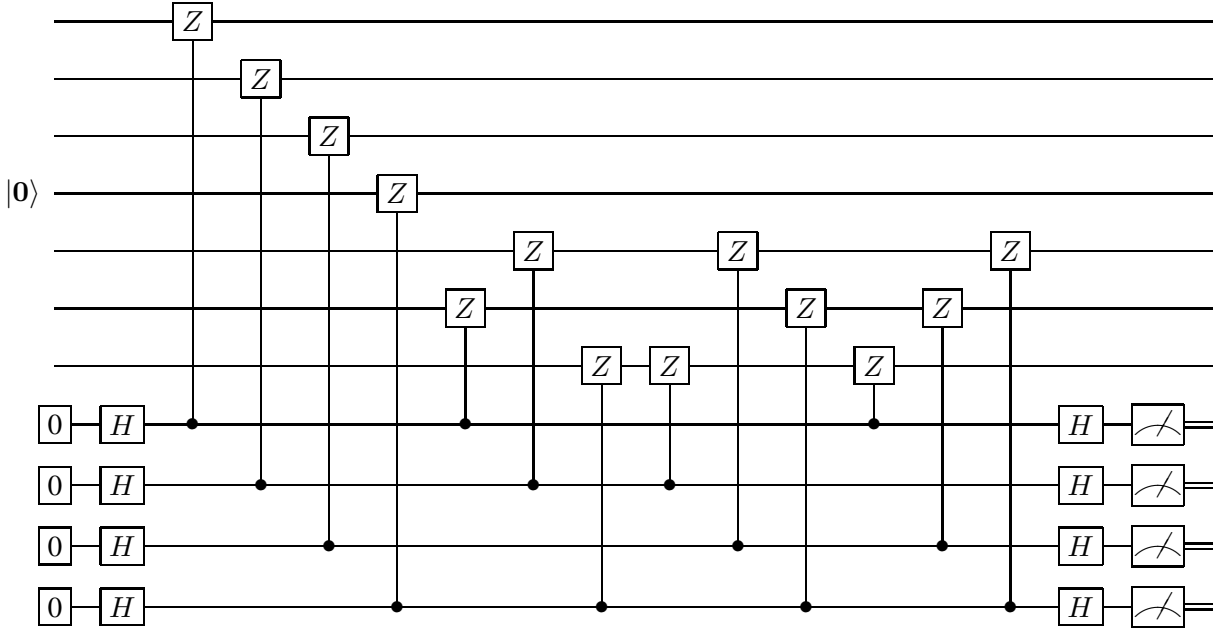
\begin{figure}[h]
\mbox{ \Qcircuit @C=1em @R=.7em {
&\qw&\gate{Z}&\qw&\qw&\qw&\qw&\qw
&\qw&\qw&\qw&\qw&\qw&\qw
&\qw&\qw&\qw&\qw \\
&\qw&\qw&\gate{Z}&\qw &\qw&\qw&\qw&\qw&\qw&\qw &\qw&\qw&\qw
&\qw&\qw&\qw&\qw&\\
&\qw&\qw&\qw&\gate{Z}&\qw&\qw&\qw&\qw &\qw&\qw&\qw
&\qw&\qw&\qw&\qw&\qw&\qw\\
\lstick{\ket{\bf 0}}&\qw&\qw&\qw &\qw&\gate{Z}&\qw&\qw &\qw&\qw
&\qw&\qw&\qw&\qw&\qw&\qw&\qw&\qw\\
&\qw&\qw &\qw&\qw&\qw &\qw&\gate{Z}&\qw&\qw&\gate{Z}&\qw&\qw&\qw
&\gate{Z}&\qw&\qw&\qw\\
&\qw&\qw&\qw &\qw&\qw&\gate{Z}&\qw&\qw&\qw&\qw&\gate{Z}&\qw&\gate{Z}&\qw
&\qw&\qw&\qw\\
&\qw&\qw&\qw&\qw&\qw&\qw&\qw&\gate{Z}&\gate{Z}&\qw&\qw&\gate{Z}&\qw&\qw
&\qw&\qw&\qw\\
\push{\fbox{0}\rule{0em}{0em}}&\gate{H}&\ctrl{-7}&\qw&\qw&\qw&\ctrl{-2}&\qw&\qw
&\qw&\qw&\qw&\ctrl{-1}&\qw&\qw&\gate{H}&\meter&\cw\\
\push{\fbox{0}\rule{0em}{0em}}&\gate{H}&\qw&\ctrl{-7}&\qw&\qw&\qw&\ctrl{-4}&\qw
&\ctrl{-2}&\qw&\qw&\qw&\qw&\qw&\gate{H}&\meter&\cw\\
\push{\fbox{0}\rule{0em}{0em}}&\gate{H}&\qw&\qw&\ctrl{-7}&\qw&\qw&\qw&\qw&\qw&\ctrl{-5}&\qw&\qw
&\ctrl{-4}&\qw&\gate{H}&\meter&\cw\\
\push{\fbox{0}\rule{0em}{0em}}& \gate{H}&\qw&\qw&\qw&\ctrl{-7}
&\qw&\qw&\ctrl{-4}&\qw&\qw&\ctrl{-5}&\qw&\qw&\ctrl{-6}
&\gate{H}&\meter&\cw } } \caption{The $\mathcal{V}$ network
\protect{\cite{preskill:faulttol,steane:fast}}, executable in 6
time steps. The boxed zero represents preparation of the
$|0\rangle$ state.  The state $\ket{\bf 0}$ is the seven-qubit
encoded $\ket{0}$ state.  If each measurement output is 0, then
the ancilla block is deemed `good', that is, it has been checked
for $X$ errors.  The network is the same for the $Z$-error
correction procedure.} \label{V_parallel}
\end{figure}

\begin{figure}[ht]
\mbox{ \Qcircuit @C=1em @R=1em {
&\gate{Z}&\qw&\qw&\qw&\qw&\qw&\qw&\qw&\qw\\
&\qw&\gate{Z}&\qw&\qw&\qw&\qw&\qw&\qw&\qw\\
&\qw&\qw&\gate{Z}&\qw&\qw&\qw&\qw&\qw&\qw\\
\lstick{\mbox{Data}}&\qw&\qw&\qw&\gate{Z}&\qw&\qw&\qw&\qw&\qw\\
&\qw&\qw&\qw&\qw&\gate{Z}&\qw&\qw&\qw&\qw\\
&\qw&\qw&\qw&\qw&\qw&\gate{Z}&\qw&\qw&\qw\\
&\qw&\qw&\qw&\qw&\qw&\qw&\gate{Z}&\qw&\qw\\
&\ctrl{-7}&\qw&\qw&\qw&\qw&\qw&\qw&\gate{H}&\meter&\cw&&&\\
&\qw&\ctrl{-7}&\qw&\qw&\qw&\qw&\qw&\gate{H}&\meter&\cw&&&\\
&\qw&\qw&\ctrl{-7}&\qw&\qw&\qw&\qw&\gate{H}&\meter&\cw&&&\\
\lstick{\mbox{Ancilla}}&\qw&\qw&\qw&\ctrl{-7}&\qw&\qw&\qw&\gate{H}&\meter&\cw&\push{\mbox{EE}}\gategroup{8}{11}{14}{13}{.5em}{=}&&\\
&\qw&\qw&\qw&\qw&\ctrl{-7}&\qw&\qw&\gate{H}&\meter&\cw&&&\\
&\qw&\qw&\qw&\qw&\qw&\ctrl{-7}&\qw&\gate{H}&\meter&\cw&&&\\
&\qw&\qw&\qw&\qw&\qw&\qw&\ctrl{-7}&\gate{H}&\meter&\cw&&& } }
\caption{The syndrome network $\mathcal{S}$ for $X$-error
correction \protect{\cite{preskill:faulttol}}. This network can be
executed in 3 time steps. Here EE represents classical error
extraction. The network $\mathcal S$ for $Z$-error correction uses
$C^X$ gates in place of $C^Z$ gates, with the ancillas acting as
control and the data as target qubits.} \label{S}
\end{figure}
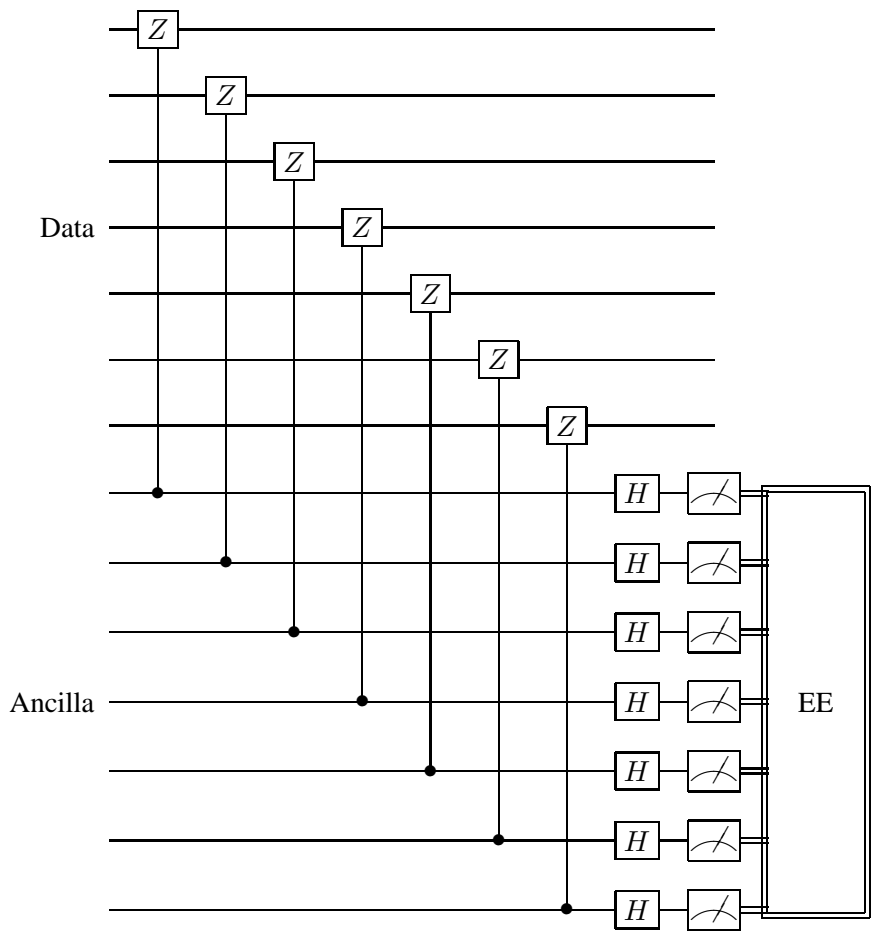

\newpage

\section{Gate Counts}\label{gatecounts}

We calculate the number of locations in the circuits $\mathcal G$, $\mathcal V$, $\mathcal S$,
 and $\mathcal R$ for the recovery gate (see Figs.\ \ref{G_parallel}, \ref{V_parallel},\ref{S}).
 Note that when recovery takes place, a one-qubit gate is executed on the data.
 We denote these numbers as $N(i \in \mathcal{G})$, etc.


\begin{table}[h]
\begin{tabular}{|c|c|c|c|c|c|c}
  &  $1$ & $2$ & $w1$ & $w2$ & $1m$ & $p$ \\ \hline\hline
$\mathcal{S}$ &  $0$ & $7$  & $14$ (on data) & $0$ & $7$ & $0$ \\
\hline $\mathcal{G}$ & $3$ & $9$  & $4$ & $3$ & $0$ & $7$ \\
\hline $\mathcal{V}$ & $4$ & $13$ & $14$ & $15+3$ & $4$ &
$4$ \\ \hline $\mathcal{R}$ & $1$ & $0$ & $6$ & $0$ & $0$ & $0$\\
\hline
\end{tabular}
\caption{Number of locations of each type ($1$, $2$, $w1$, $w2$, $1m$, or $p$) in individual routines $\mathcal G$, $\mathcal V$, $\mathcal S$ and the recovery gates $\mathcal R$. The $w1$ and $w2$ locations combined are simply called $w$ locations in the nonlocal analysis.}
\label{locGVSR}
\end{table}

\bibliographystyle{hunsrt}


\end{document}